\documentclass[pra,twocolumn,showpacs,aps]{revtex4-1}
\usepackage{color}
\usepackage{amsmath,amssymb}
\usepackage{txfonts}
\usepackage{graphicx}

\usepackage[unicode]{hyperref}
\hypersetup{
    pdftitle={Low-dimensional stochastic projected Gross-Pitaevskii equation},
    pdfauthor={A. S. Bradley, S. J. Rooney and R. G. McDonald},
    pdfsubject={Low-D SPGPE},
    colorlinks=true,
    linkcolor=blue,
    citecolor=blue,
    filecolor=black,
    urlcolor=blue
}
\usepackage{bm}
\usepackage{epstopdf}
\usepackage{enumerate}
\def\urlprefix{}
   \def\url#1{}

\newcommand{\ve}{\varepsilon}

\newcommand{\ec}{\epsilon_{\textrm{cut}}}
\newcommand{\rC}{C}
\newcommand{\rI}{I}

\newcommand{\ket}[1]{|#1\rangle}
\newcommand{\bra}[1]{\langle#1|}

\newcommand{\EQ}[1]{\begin{align}#1\end{align}}

\newcommand{\mbf}[1]{\mathbf{#1}}

\newcommand{\Rb}{$^{87}$Rb}
\newcommand{\QQ}{{\cal Q}}
\newcommand{\hqq}{\hat{\cal Q}}
\newcommand{\PP}{{\cal P}}
\newcommand{\hpp}{\hat{\cal P}}
\newcommand{\LL}{L}

\newcommand{\kk}{\mbf{k}}

\newcommand{\rr}{\mathbf{r}}
\newcommand{\rrr}{\mathbf{r}_\perp}

\newcommand{\eref}[1]{(\ref{#1})}
\newcommand{\fref}[1]{Fig.~\ref{#1}}

\begin{document}
\title{Low-Dimensional Stochastic Projected Gross-Pitaevskii Equation}
\author{A.~S. Bradley, S. J. Rooney and R. G. McDonald} 
\affiliation{Department of Physics, QSO | Centre for Quantum Science, and Dodd-Walls Centre for Photonic and Quantum Technologies, University of Otago, Dunedin 9010, New Zealand.}
\begin{abstract}
We present reduced-dimensional stochastic projected Gross-Pitaevskii equations describing regimes of confinement and temperature where a 1D or 2D superfluid is immersed in a 3D thermal cloud. The projection formalism provides both a formally rigorous and physically natural way to effect the dimensional reduction. The 3D form of the number-damping (growth) terms is unchanged by the dimensional reduction. Projection of the energy-damping (scattering) terms leads to modified stochastic equations of motion describing energy exchange with the thermal reservoir. The regime of validity of the dimensional reduction is investigated via variational analysis. Paying particular attention to 1D, we validate our variational treatment by comparing numerical simulations of a trapped oblate system in 3D with the 1D theory, and establish a consistent choice of cutoff for the 1D theory. We briefly discuss the scenario involving two-components with different degeneracy, suggesting that a wider regime of validity exists for systems in contact with a buffer-gas reservoir. 
\end{abstract}
\maketitle
\section{Introduction}
The stochastic projected Gross-Pitaevskii equation (SPGPE)~\cite{Gardiner:2003bk} was derived via a rigorous treatment of the reservoir interaction process governing the evolution of high-temperature Bose-Einstein condensates (BECs). The theory emerged as a synthesis of quantum kinetic theory~\cite{QKI,QKIII,Gardiner:2000fi} and the projected Gross-Pitaevskii equation (PGPE)~\cite{Davis2001b,Davis2002,Blakie05a}. The former gives a treatment of condensate growth in experiments from a particle kinetics point of view~\cite{QKPRLII,QKPRLIII}, and the latter generalizes the Gross-Pitaeveskii equation to describe systems where the condensate is one of many degenerate modes | precisely the physical situation near the BEC transition~\cite{Davis:2006ic,Blakie2007a,Bezett09b,Bezett09a}. The central principle allowing unification of the two approaches is the use of a \emph{projector} to separate the complete system into subsystems that may be treated in different levels of approximation. The projector implements a high-energy cutoff, enabling a dynamical description of  the low-energy system of interest, both formally~\cite{Gardiner:2003bk} and numerically~\cite{Bradley:2008gq,Rooney:2010dp,Rooney:2011fm}, in the Wigner phase-space representation. This general procedure is familiar from the open systems theory of quantum optics~\cite{QN,QO}, and leads to a stochastic differential equation governing the system evolution containing two kinds of dissipative terms: those associated with collisions causing particle exchange with the reservoir (growth/loss), and those causing damping and diffusion of \emph{energy} without particle transfer~\cite{Rooney:2012gb}. The simplified SPGPE derived by neglecting the energy damping terms (the simple growth SPGPE~) has been used to study the growth of BEC from a rotating thermal cloud~\cite{Bradley:2008gq}, the formation of spontaneous vortices via the Kibble-Zurek mechanism~\cite{Weiler:2008eu}, the decay of a vortex~\cite{Rooney:2010dp,Rooney:2011fm}, to the formation of a persistent current~\cite{Rooney:2013ff}, to the onset of quasicondensation in an elongated 3D system~\cite{Garrett:2013gk}, and to modelling thermal fluctuations in a continuously pumped atom laser~\cite{Lee:2015vd}. In application to thermometry in 1D, this description gave excellent agreement with exact Yang-Yang thermomety~\cite{Davis:2012hq}.

The energy damping collisions between low energy atoms and reservoir atoms are a source of phase noise~\cite{Anglin:1997cf,Anglin:1999fn,TrujilloMartinez:2009gh,Poletti:2012di,Kordas:2013gw}, stem from the \emph{quantum brownian motion} master equation~\cite{QN},
and cause significant damping process for highly non-equilibrium dynamics~\cite{Blakie:2008is,Rooney:2012gb}. Even in quasi-static systems, the energy damping process can set up superfluid counterflow if the reservoir develops a temperature gradient~\cite{Gilz:2011jma}. 
Work on the full SPGPE that includes these terms has focused on numerically implementing the energy damping (scattering) terms in the SPGPE~\cite{Rooney:2012gb,Rooney:2014kc}, and on generalising the theory to spinor and multicomponent systems~\cite{Bradley:2014a}. Despite these developments, there remains much to be understood about the theory and its implications. In particular, a complete picture of the dissipative dynamics of excitations in BEC such as vortices and solitons~\cite{Cockburn10a,Cockburn:2011fa,Rooney:2010dp} requires a deeper understanding of the energy-damping terms. Energy damping may also play a role in the Kibble-Zurek mechanism of spontaneous vortex formation during the BEC transition~\cite{Weiler:2008eu,Su:2013dh,Damski10a,Navon:2015jd}, and in 2D quantum turbulence (2DQT) where many vortices confined to planar motion interact to collectively transport energy to large scales~\cite{Bradley:2012ih,Reeves:2013hy,Neely:2013ef}. 
\par
Solving the three-dimensional SPGPE \cite{Rooney:2012gb} is in general a numerically challenging task, due to the need for an exact energy cutoff (requiring a single-particle basis for numerical propagation~\cite{Bradley:2008gq}), and the additional expense of evaluating the non-local energy-damping term and integrating a stochastic differential equation containing multiplicative noise~\cite{Rooney:2014kc}. However, for systems that are tightly constrained in one or two dimensions, the implementation of the energy cutoff via a projector provides a natural route to obtaining an effective theory in the low-dimensional subspace of weakly confined dimensions. Experimentally, a high level of control over confinement geometry is accessible optically~\cite{Arnold:2011gy,Henderson:2009eo,Gaunt:2013ip}, allowing the study of dissipative superfluid phenomena in new settings. 
\par
In this work we consider systems with strong confinement along one or two spatial dimensions and project the 3D-SPGPE onto the ground state of the harmonic trap in each of the dimensions of strong confinement, to obtain an effective SPGPE in the subspace to which dynamics are confined. This procedure involves choosing the energy cutoff so as to eliminate all transverse degrees of freedom. Identifying the regime of validity of this procedure requires some care, as the reservoir interaction theory we use requires that the system has a 3D thermal reservoir; if the reservoir is also low-dimensional, the collision integrals determining the reservoir interaction rates become nontrivially spatially dependent~\cite{Gardiner:2003bk,Bradley:2008gq,Bradley:2014a}. We assess the validity of dimensional reduction using variational analysis, and for the 1D regime, by comparison with 3D numerical simulations. 

\section{High temperature C-field theory}
The \emph{projected} open systems theory of high-temperature BEC begins by setting up an energy cutoff in an appropriate single-particle basis of modes representing the system. Where possible the basis is chosen to  diagonalise all time-independent linear terms in the Hamiltonian, since it is then a good basis for imposing an energy cutoff for the interacting system (provided the cutoff is chosen sufficiently high). 
\par 
The system of interacting bosons confined by external trapping potential $V_T(\rr,t)=V_0(\rr)+\delta V(\rr,t)$ is described by the Hamiltonian 
\EQ{\label{Hfull}
H=&{}\int d^3\rr\;\hat\Psi^\dag(\rr)\left[{\cal H}(\rr)+\delta V(\rr,t)\right]\hat\Psi(\rr)\nonumber\\
&{}+\frac{u}{2}\int d^3\rr\;\hat\Psi^\dag(\rr)\hat\Psi^\dag(\rr)\hat\Psi(\rr)\hat\Psi(\rr)
}
where the time-independent single-particle Hamiltonian density is
\EQ{
{\cal H}(\rr)&=-\frac{\hbar^2\nabla^2}{2m}+ V_{0}(\rr ),
}
and $u=4\pi\hbar^2 a/m$ is the interaction parameter for s-wave scattering length $a$. The field operators satisfy the usual equal-time Bose commutation relations, with non-vanishing commutator
\EQ{\label{bcom}
[\hat\Psi(\rr),\hat\Psi^\dag(\rr')]=\delta(\rr-\rr').
}
The system is represented in terms of the single-particle eigenfunctions $\phi_n(\rr)$, that satisfy
\EQ{\label{speigs}
{\cal H}(\rr)\phi_n(\rr)=\epsilon_n\phi_n(\rr),
}
where the index $n$ denotes all quantum numbers defining a unique eigenfunction. The field operator expansion then reads
\EQ{\label{fop}
\hat\Psi(\rr)&=\sum_n\hat a_n\phi_n(\rr),
}
for single-mode operators $\hat a_n$ satisfying $[\hat a_n,\hat a_m^\dag]=\delta_{nm}$.
Our first task is to consistently separate the system into a $C$-region, where populations are appreciable and the atoms are at least partially coherent, and an $I$-region where populations are low and atoms are incoherent. We then seek an equation of motion for the $C$-region, treating the $I$-region as an incoherent reservoir. A significant advantage of effecting the separation in the single-particle basis is that at sufficiently high energy it diagonalises the many-body problem, thus providing a good basis for separating the system, provided the separation energy is large. We define the $C$-region as $C=\{\epsilon_n\leq\epsilon_{\textrm{cut}}\}$, where $\ec$ will be significantly larger than the system chemical potential $\mu$ (of order $2\mu-3\mu$). In this basis we introduce the orthogonal projection operators 
\EQ{\label{Pdef}
\hpp&\equiv\sum_{\epsilon_n\leq\ec}\ket{n}\bra{n},\\
\hqq&\equiv 1-\hpp,
}
satisfying $\hpp\hpp=\hpp$, $\hqq\hqq=\hqq$, $\hqq\hpp=0$. In the position representation the field operator decomposes into $\rC$-region and $I$-region operators as
\EQ{\label{cfield}
\hat\Psi(\rr)&= \PP\hat\Psi(\rr)+\QQ\hat\Psi(\rr)\equiv \hat\psi(\rr)+\hat\eta(\rr)
}
respectively, where 
\EQ{\label{Px}
\hat\psi(\rr)&= \PP\hat\Psi(\rr)\equiv\sum_{\epsilon_n\leq \ec}\phi_n(\rr)\int d^3\rr\;\phi^*_n(\rr)\hat \Psi(\rr)\\
&=\sum_{\epsilon_n\leq \ec}\hat a_n\phi_n(\rr),
}
defines the spatial representation of $\hpp$ and the projected field operator with commutator
\EQ{\label{dC}
[\hat\psi(\rr),\hat\psi^\dag(\rr')]=\delta(\rr,\rr')\equiv \sum_{\epsilon_n\leq \ec}\phi_n(\rr)\phi^*_n(\rr').
}
This formal separation of the system provides a natural approach to deriving an equation of motion describing the evolution of the $\rC$-region, the details of which can be found elsewhere~\cite{Gardiner:2003bk,Bradley:2014a}. The derivation proceeds by mapping the master equation for the $\rC$-region density operator to an equation for the Wigner distribution of the system; an equivalent diffusion process for a classical field $\psi(\rr)$ (c-field) is then obtained, the moments of which correspond to symmetrically ordered averages of the field operator at equal times~\cite{Steel:1998jr,Blakie:2008is}. The $I$-region field $\hat\eta(\rr)=\QQ\hat\Psi(\rr)$ enters the equation through the rate functions determining the strength of reservoir interaction processes~\cite{Bradley:2008gq,Rooney:2012gb}. 
\par
For our purposes, we take as our starting point the three-dimensional Stratonovich SPGPE for the $\rC$-field~\cite{Rooney:2012gb}. Taking our energy reference as $\mu$, the SPGPE takes the form
\EQ{\label{3dSPGPE}
(S) d\psi(\rr,t)&=d\psi\Big|_H+d\psi\Big|_\gamma+(S)d\psi\Big|_\ve,
}
with
\EQ{
\label{spgpeH}
i\hbar d\psi\Big|_H&=\PP\left\{\LL\psi dt\right\},\\
\label{spgpeG}
i\hbar d\psi\Big|_\gamma&=\PP\left\{-i\gamma\LL\psi dt+i\hbar dW(\rr,t)\right\},\\
\label{spgpeE}
(S)i\hbar d\psi\Big|_\ve&=\PP\left\{V(\rr,t)\psi dt-\hbar\psi dU(\rr,t)\right\}.
}
The Hamiltonian evolution relative to the reservoir chemical potential is generated by
\EQ{\label{Ldef}
\LL\psi(\rr,t)&\equiv \left[{\cal H}(\rr)+\delta V(\rr,t)+u|\psi(\rr,t)|^2-\mu\right]\psi(\rr,t),
}
thus recovering the PGPE~\cite{Blakie05a}
\EQ{\label{pgpe}
i\hbar\frac{\partial \psi(\rr,t)}{\partial t}&=\PP\{ L\psi(\rr,t)\}.
}
This equation of motion describes the evolution of a low-energy fraction of atoms with partial coherence, including the condensate and many excitations~\cite{Davis2001a,Blakie05a,Wright:2011ey}\footnote{Note that in general $\delta V(\rr,t)$ in \eref{Ldef} can be redefined to include the effective potential arising from forward scattering with the $I$-region~\cite{Gardiner:2003bk,Bradley:2014a}. This term is typically a small correction and we do not consider it further here.}.
The high energy reservoir coupled to the PGPE is described by chemical potential ($\mu$), temperature ($T$), and cutoff energy ($\ec$), and the functions
\EQ{
\gamma&=\frac{8 a^2}{\lambda_{dB}^2}\sum_{j=1}^\infty \frac{e^{\beta\mu(j+1)}}{e^{2\beta\ec j}}\Phi\left[\frac{e^{\beta\mu}}{e^{2\beta\ec}},1,j\right]^2,\\
V(\rr,t)&=-\hbar \int d^3\rr^\prime \ve(\rr-\rr^\prime)\nabla^\prime\cdot\mathbf{j}(\rr^\prime,t),\\
\mathbf{j}(\rr,t)&=\frac{i\hbar}{2m}\left[\psi\nabla\psi^*-\psi^*\nabla\psi\right],\\
\ve(\rr)&=\frac{\cal M}{(2\pi)^3}\int d^3\kk \frac{e^{i\kk\cdot \rr}}{|\kk|},\\
{\cal M}&\equiv \frac{16 \pi a^2}{e^{\beta(\ec-\mu)}-1},
}
where $\lambda_{dB}=\sqrt{2\pi \hbar^2/m k_B T}$, $\beta=1/k_B T$, and $\Phi[z,x,a]=\sum_{k=0}^\infty z^k/(a+k)^x$ is the \emph{Lerch transcendent}, and where $\gamma$ and $\ve(\rr)$ are both dimensionless. The noise terms are Gaussian, with non-vanishing correlations
\EQ{
\langle dW^*(\rr,t)dW (\rr^\prime,t)\rangle=\frac{2k_B T }{\hbar}\gamma\delta(\rr',\rr) dt, \\
\langle dU(\rr,t)dU(\rr^\prime,t)\rangle=\frac{2k_B T}{\hbar}\ve(\rr-\rr^\prime) dt,
}

For a reservoir close to equilibrium, the growth rate $G(\rr)$ (See Eq. (6) of Ref. \cite{Rooney:2012gb}) acquires the same value over most of the $\rC$-field region, and varies slowly near its edges~\cite{Bradley:2008gq}, justifying our approximation of the growth in the 3D system by  its spatially independent form, and defining the growth parameter $ \hbar G(\rr)/k_B T\approx \gamma$. This approximation has the additional advantage of putting the treatment of growth and scattering on an equal footing: for a thermal cloud near equilibrium the coefficient of the scattering term is spatially invariant (aside from the irreducibly non-local scattering kernel). In \eref{3dSPGPE} and what follows we use the subscript $\ve$ ($\gamma$) to denote the the scattering (growth) reservoir interaction terms in the time-independent reservoir approximation used in this work. We emphasize that the energy cutoff defining the reservoir is taken at quite a high energy (relative to $\mu$), and the dimensionless damping rates are quite small ($\sim 10^{-4}-10^{-3}$ in experiments~\cite{Rooney:2013ff}), so that in practice this approximation allows for significant dynamics of the non-condensed atoms in the $\rC$-field.

\begin{figure}[!t]{
\begin{center} 
\includegraphics[width=\columnwidth]{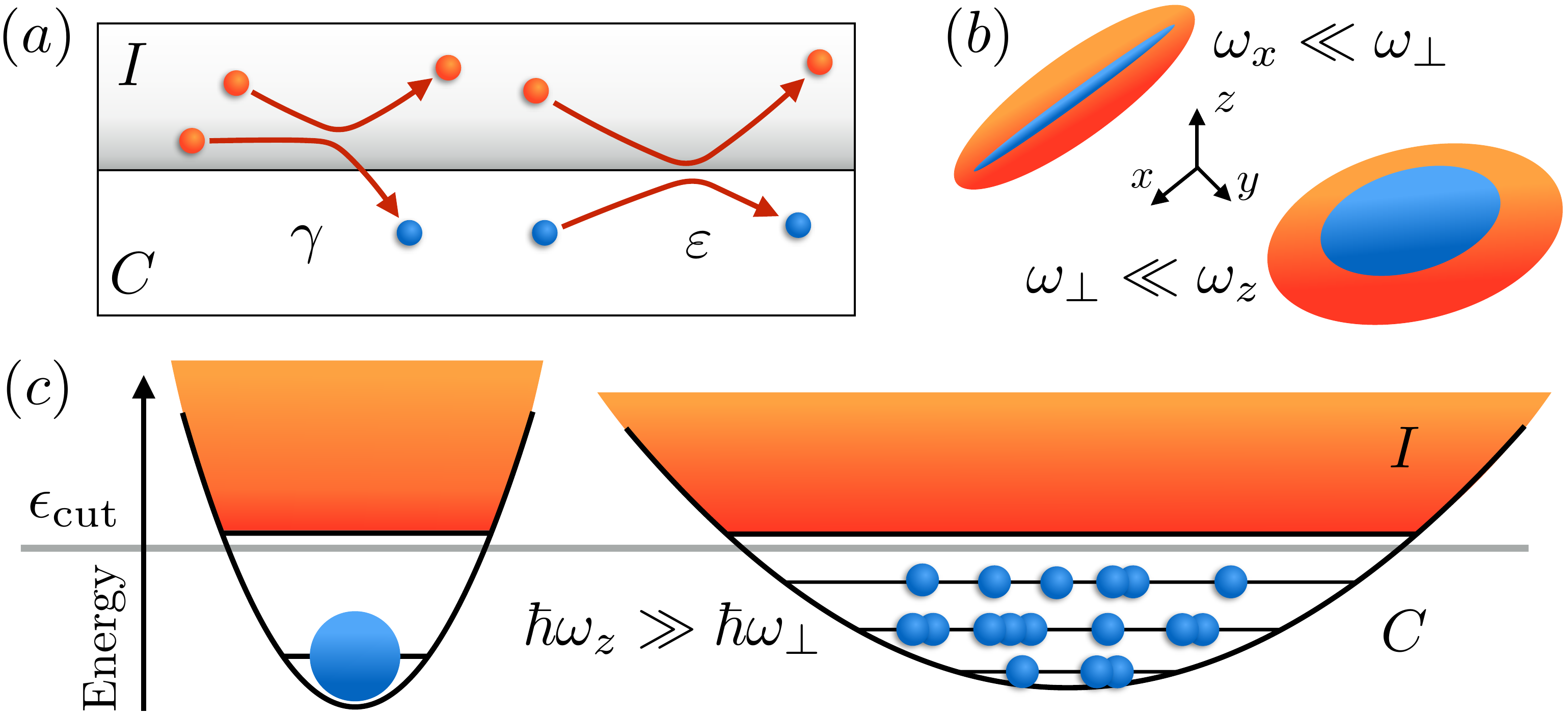}
\caption{(Color online) Schematic of Low-D SPGPE regimes for an anisotropic 3D Bose gas. (a) Reservoir interaction processes for number damping ($\gamma$) and energy damping ($\ve$) respectively. (b) 1D and 2D scenarios of anisotropy with low-D c-field and 3D thermal cloud. (c) Separation of energy scales due to anisotropy allows the transverse degrees of freedom to be treated as part of the thermal reservoir.
\label{fig1}}
\end{center}}
\end{figure}

Defining, respectively, the $\rC$-field atom number, Hamiltonian, and grand-canonical energy as
\EQ{\label{ndef}
N&=\int d^3\rr\;|\psi|^2,\\
\label{hdef}
H&=\int d^3\rr\;\psi^*\left({\cal H}+\frac{u}{2}|\psi|^2\right)\psi,\\
\label{kdef}
K&=H-\mu N,
}
we can get some insight as to the role of the terms in the equation of motion.
Without noise, i.e. formally setting all noises to zero in the SPGPE, the equations of motion generate the dissipative evolution
\EQ{
\frac{dN}{dt}=-\frac{2\gamma}{\hbar}(\mu(t)-\mu)N,\\
} 
where $\mu(t)\equiv\int d^3\rr\;\psi^*(\LL+\mu)\psi/N$ is the instantaneous chemical potential (the definition \eref{Ldef} introduces a shift by $\mu$), and
\EQ{
\frac{dK}{dt}=-\frac{2\gamma}{\hbar}\int d^3\rr\;|\LL\psi|^2-\hbar{\cal M}\int \frac{d^3\kk}{|\kk|}|\kk\cdot \mathbf{j}(\kk)|^2
}
is non-negative.
Clearly $\gamma$ describes \emph{number damping} (with associated energy damping), while $\ve$ describes \emph{energy damping} (without associated number damping), and both processes arise from scattering between $\rC$-region and $\rI$-region atoms, with rates that are quadratic in the scattering length. The two collision processes responsible for stochastic evolution of the SPGPE are shown schematically in \fref{fig1} (a).
\section{Dimensional reduction}
To introduce notation, we first consider the reduction to a 2D subspace, i.e. a system with tight parabolic confinement along the $z$-axis, defined by harmonic oscillator frequency $\omega_z$. Defining $\rr_\perp=(x,y)$, we then have
\EQ{\label{VT1}
V_0(\rr)\equiv V_0(\rr_\perp)+\frac{m\omega_z^2 z^2}{2}.
}
We can ensure that the system has a 3D thermal cloud by imposing the temperature condition
\EQ{\label{2dsubsys}
\hbar\omega_z\lesssim k_BT.
}

Then it will be a good approximation to introduce the energy cutoff as $\ec\sim\hbar\omega_z/2$, i.e. we treat modes above the ground state in the $z$-dimension as part of the $\rI$-region. Since the energy scale in the $z$-direction is quite large, the mode number in the 2D subspace can still be varied, as $\ec$ can be chosen in the range $\hbar\omega_z/2\leq \ec<3\hbar\omega_z/2$ while still maintaining a formal dimensional reduction. The dynamics of the superfluid will remain confined to the plane provided the 2D chemical potential satisfies $\mu_2\ll\hbar\omega_z$. The validity condition for dimensional reduction can be expressed as
\EQ{\label{2Dcond}
\mu_2\ll \hbar\omega_z\lesssim k_BT.
}
\par
The situation is very similar in 1D. The trapping potential is now assumed to have strong parabolic confinement along the $y$- and $z$-axes:
\EQ{\label{VT1}
V_0(\rr)\equiv V_0(x)+\frac{m\omega_\perp^2 (y^2+z^2)}{2}.
}
Putting the two conditions together, a consistent 3D reservoir theory of the 1D subsystem is obtained provided the condition 
\EQ{\label{1Dcond}
\mu_1\ll\hbar\omega_\perp\lesssim k_BT
}
is satisfied.  It will be helpful in what follows to define the $D$-dimensional transverse frequency as
\EQ{\label{OmD}
\Omega_D=\begin{cases}
\omega_z& (D=2),\\
\omega_\perp & (D=1),
\end{cases}
}
in terms of which our validity condition reads
\EQ{\label{OmVal}
\mu_D\ll\hbar\Omega_D\lesssim k_BT.
}
These two scenarios of anisotropy are shown schematically in \fref{fig1} (b); the separation of energy scales is illustrated for the pancake configuration in \fref{fig1} (c). 
The first inequality in \eref{OmVal} raises a central question of this paper: how different must these scales be for validity to be ensured? We will return to this question in Secs.~\ref{sec:varVal} and \ref{1dval}.
In what follows we will also find it convenient to introduce the oscillator length scale associated with the transverse trapping potential:
\EQ{\label{sigD}
\sigma_D&=\sqrt{\frac{\hbar}{m\Omega_D}}.
}
To keep notation simple, we will often refer simply to $\Omega$ and $\sigma$ as it will be clear from the context which quantity is being used. For parabolic traps it also customary to define the oscillator lengths $a_x=\sqrt{\hbar/m\omega_x}$, etc, and these quantities will be reverted to where appropriate. 

In each of the tightly confined dimensions we insist that the $\rC$-field is separable, and that the system is in the transverse ground state wave function, e.g.
\EQ{\label{gnd}
\phi_0(z)=\left(\frac{1}{\pi\sigma^2}\right)^{1/4}e^{-z^2/2\sigma^2},
}
where the subscript of $\sigma_D$ will be omitted unless it is essential for clarity. Dimensional reduction is carried out by integrating over the tightly confined dimensions. We can state this procedure formally noting that we can always (since our basis is separable) write the projector in the factorized form
\EQ{\label{facP}
\PP&\equiv \PP_D\PP_{3-D},
}
where $\PP_D$ is the reduced dimensional projector for the $D$-dimensional subspace [the low-D form of \eref{Px}], and $\PP_{3-D}$ is the projection onto the transverse degrees of freedom; for the regime we consider, the transverse projector can be evaluated explicitly as the projection onto \eref{gnd} in each dimension.
\par
To avoid overly cumbersome notation we denote the dimensionality of the wavefunction by its spatial argument $\rr$, $\rrr$, and $x$ in 3D, 2D and 1D respectively.
\subsection{Number damping}
Since the number damping rate $\gamma$ is well approximated as spatially invariant, the number damping terms in low-D are formally identical to their 3D form. Dimensional reduction leads to a modified interaction strength, as has been described in many previous works on the GPE~\cite{Steel:1998jr}. The dimensionally reduced forms for $N$, $H$, $K$ may be summarized as
\EQ{\label{ND}
N_D&=\int d^D\rr\;|\psi |^2,\\
H_D&=\int d^D\rr\;\psi^*\left({\cal H}_D+\frac{g_D}{2}|\psi|^2\right)\psi,\\
K_D&=H_D-\mu_D N_D,
}
where the interaction parameter
\EQ{\label{gD}
g_D&=
\begin{cases}
4\pi\hbar^2 a/m & (D=3),\\
\sqrt{8\pi}\hbar^2a/m\sigma & (D=2),\\
2\hbar\Omega a,& (D=1),\\
\end{cases}
}
is obtained by integrating out the transverse ground-state wavefunction in 2D and 1D respectively, and the chemical potential is given by
\EQ{\label{muD}
\mu_D&=
\begin{cases}
\mu & (D=3),\\
\mu-\hbar\omega_z/2 & (D=2),\\
\mu-\hbar\omega_\perp& (D=1).\\
\end{cases}
}
The single-particle Hamiltonian density is
\EQ{\label{HspD}
{\cal H}_D&=
\begin{cases}
{\cal H}& (D=3),\\
-\hbar^2\nabla_\perp^2/2m+V_T(\rr_\perp,t)& (D=2),\\
-\hbar^2\partial_x^2/2m+V_T(x,t)& (D=1).\\
\end{cases}
}
The dimensional reduction applied to \eref{spgpeH} and \eref{spgpeG} then gives
\EQ{\label{ndampnd}
i\hbar d\psi&=\PP_D\left\{(1-i\gamma)\LL_D\psi dt+i\hbar dW_D(\rr,t)\right\},
}
where
\EQ{\label{LD}
L_D\psi&\equiv ({\cal H}_D+g_D|\psi|^2-\mu_D)\psi,
}
\EQ{\label{dWD}
\langle dW_D^*(\rr,t)dW_D (\rr',t)\rangle&=\frac{2k_B T }{\hbar}\gamma\delta_D(\rr',\rr) dt, 
}
$\delta_D(\rr',\rr)$ is the $D$-dimensional projected delta function, and $\rr$ takes the form appropriate for subspace $D$. When convenient, we will also refer to these projectors as $\PP_x$ and $\PP_\perp$ in 1D and 2D respectively. Thus, in the regime where the thermal cloud remains 3D (i.e. well away from the extreme low-D limit where S-wave scattering is modified), the dimensional reduction for the number-damping terms yields only a renormalized interaction strength, and a new reference energy for the chemical potential.  
\subsection{Energy damping in 1D}
We will describe one dimension derivation in detail to give an indication of the procedure for formally obtaining the equation of motion for the theory in a reduced dimensional subspace. 
The energy damping terms are found by projecting the full field $\psi(\rr,t)=\phi_0(y)\phi_0(z)\psi(x,t)$ as
\EQ{\label{SPsi}
(S)i\hbar d\psi(x,t)\Big|_\ve={}&\int dy\int dz\;\phi_0^*(y)\phi_0^*(z){\cal P}_x\Bigg\{\nonumber\\
{}& V(\rr,t)\psi(\rr,t)dt\nonumber\\
{}&-\hbar\psi(\rr,t)dU(\rr,t)\Bigg\}.
}
This gives the one-dimensional form 
\EQ{\label{Spsi}
(S)i\hbar d\psi(x,t)\Big|_\ve={}&{\cal P}_x\Bigg\{V_1(x,t)\psi(x,t)dt\nonumber\\
{}&-\hbar \psi(x,t)dU_1(x,t)\Bigg\}
}
with new potential and noise obtained by evaluating the projectors for each term. We thus define 
\EQ{
V_1(x,t)\equiv{}& \int dy\int dz\;|\phi_0(y)|^2|\phi_0(z)|^2V(\rr,t)\nonumber\\
\label{1Deps}
={}&-\hbar\int dx'\ve_1(x-x')\partial_{x'}j(x',t),
}
in terms of the function
\EQ{\label{1Depseval}
\ve_1(x-x')={}&\frac{{\cal M}}{(2\pi)^3}\int dk_x\;e^{ik_x(x-x')}\int \frac{d^2\kk_\perp}{|\kk|}\nonumber\\
{}&\times\int d^2\rr_\perp'\int d^2\rr_\perp\;e^{i\kk_\perp\cdot(\rr_\perp-\rr_\perp')}\frac{e^{-r_\perp^2/\sigma^2}}{\pi\sigma^2}\frac{e^{-r_\perp'^2/\sigma^2}}{\pi\sigma^2}\nonumber\\
={}&\frac{{\cal M}}{(2\pi)^2\sigma}\int dk_x e^{ik_x(x-x')}\int_0^\infty \frac{du\; u e^{-u^2/2}}{\sqrt{u^2+(k_x\sigma)^2}},
}
where we have changed variables to $u=k_\perp \sigma$. Evaluating the integral, we obtain
\EQ{\label{1Dget}
\ve_1(x)&=\frac{{\cal M}}{2\pi}\int_{-\infty}^\infty dk\; e^{ikx}S_1(k),
}
where 
\EQ{\label{S1def}
S_1(k)&\equiv\frac{1}{\sqrt{8\pi\sigma^2}}G\left(\frac{|k|\sigma}{\sqrt{2}}\right),
}
and $G(q)\equiv e^{q^2}{\rm erfc}(q)$ is the \emph{scaled complementary error function}. 
\par
Projecting the noise onto the 1D subspace, we have
\EQ{\label{dWe1d}
dU_1(x,t)&\equiv\int dy\int dz\;|\phi_0(y)|^2|\phi_0(z)|^2dU(\rr,t).
}
Since the noises are completely determined by their first and second moments, we can instead evaluate the correlation function
\EQ{\label{dWe1d2}
\langle dU_1(x,t)dU_1(x',t)\rangle={}&\int dy\int dz\; |\phi_0(y)|^2|\phi_0(z)|^2\nonumber\\
{}&\times\int dy'\int dz' |\phi_0(y')|^2|\phi_0(z')|^2\nonumber\\
{}&\times\langle dU(\rr,t)dU(\rr^\prime,t)\rangle,
}
and evaluating the integrals gives an expression for the noise correlator in terms of \eref{1Dget} as
\EQ{\label{dWe1}
\langle dU_1(x,t)dU_1(x',t)\rangle=\frac{2k_B T}{\hbar}\ve_1(x-x^\prime)dt.
}
This reduced correlation function gives a complete description of the noise for the 1D subspace.

\subsection{Energy damping in 2D}
The energy damping terms are projected from the full field $\psi(\rr,t)=\psi(\rrr,t)\phi_0(z)$ as
\EQ{\label{SPsi}
(S)i\hbar d\psi(\rrr,t)\Big|_\ve={}&\int dz\;\phi_0^*(z){\cal P}_{xy}\Bigg\{\nonumber\\
{}& V(\rr,t)\psi(\rr,t)dt\nonumber\\
{}&-\hbar\psi(\rr,t)dU(\rr,t)\Bigg\}
}
giving the two-dimensional form
\EQ{\label{Spsi}
(S)i\hbar d\psi(\rr_\perp,t)\Big|_\ve={}&{\cal P}_\perp\Bigg\{V_2(\rr_\perp,t)\psi(\rr_\perp,t)dt\nonumber\\
{}&-\hbar\psi(\rr_\perp,t)dU_2(\rr_\perp,t)\Bigg\}
}
by evaluating the projectors on each term. We thus define
\EQ{\label{Vmx}
V_2(\rrr,t)&\equiv \int dz\; |\phi_0(z)|^2V (\rr,t)\nonumber\\
&=-\hbar\int d^2\rr'_\perp\ve_2(\rr_\perp-\rr'_\perp)\nabla_\perp'\cdot \mathbf{j}_\perp(\rr_\perp',t),
}
where
\EQ{\label{eps2}
\ve_2(\rr_\perp-\rr_\perp')={}&\frac{{\cal M}}{(2\pi)^3}\int d^2 \kk_\perp e^{i\kk_\perp\cdot(\rr_\perp-\rr_\perp')}\int_{-\infty}^\infty dk_z \frac{1}{|\kk|}\nonumber\\
{}&\times \int dz\int dz' \;e^{ik_z(z-z')}\frac{e^{-z^2/\sigma^2}}{\sqrt{\pi\sigma^2}}\frac{e^{-z'^2/\sigma^2}}{\sqrt{\pi\sigma^2}}\nonumber\\
={}&\frac{{\cal M}}{(2\pi)^3}\int d^2 \kk_\perp e^{i\kk_\perp\cdot(\rr_\perp-\rr_\perp')}\nonumber\\
{}&\times\int_{-\infty}^\infty \frac{du\;e^{-u^2/2}}{\sqrt{u^2+(k_\perp\sigma)^2}}.
}
Evaluating the integral, we arrive at the 2D result
\EQ{\label{Vmreduce}
\ve_2(\rr_\perp-\rr_\perp',t)&=\frac{{\cal M}}{(2\pi)^2}\int d^2 \kk_\perp e^{i\kk_\perp\cdot(\rr_\perp-\rr_\perp')}S_2(\kk_\perp),
}
where 
\EQ{\label{fdef}
S_2(\kk)\equiv\frac{1}{ 2\pi }F\left(\frac{(k\sigma)^2}{4}\right),
}
and $F(x)\equiv e^{x}K_0(x)$ is the \emph{scaled modified Bessel function}. 
\par
The noise is
\EQ{\label{dWx}
dU_2(\rrr,t)\equiv \int dz\;|\phi_0(z)|^2dU(\rr,t),
}
and as above, we evaluate the noise correlation in 2D
\EQ{\label{dW_M2}
\langle dU_2(\rrr,t)dU_2(\rrr^\prime,t)\rangle={}&\int dz\;|\phi_0(z)|^2\int dz^\prime |\phi_0(z^\prime)|^2\nonumber\\
{}&\times\langle dU(\rr,t)dU(\rr^\prime,t)\rangle,
}
finding 
\EQ{\label{dWM2D}
\langle dU_2(\rrr,t)dU_2(\rrr^\prime,t)\rangle=\frac{2k_B T}{\hbar}\ve_2(\rrr-\rrr^\prime)dt.
}
\subsection{Summary of Low-D SPGPE}
In this subsection we summarize the SPGPE in each of the three regimes. To simplify notation hereafter we drop the cumbersome subscript $\rr_\perp$, as it will be clear form the context that the wave function, spatial coordinate, and wave vector coordinates are set by the dimension $D$.
We define the Fourier-space scattering kernel
\EQ{\label{Sgen}
S_D(\mathbf{k})\equiv
\begin{cases}
|\mathbf{k}|^{-1}  & (D=3),\\
\frac{1}{ 2\pi}F\left(\frac{|\kk|^2\sigma^2}{4}\right) & (D=2),\\
\frac{1}{\sqrt{8\pi\sigma^2}}G\left(\frac{|k|\sigma}{\sqrt{2}}\right)& (D=1),\\
\end{cases}
}

The asymptotic scaling of $S_D(\mathbf{k})$ is
\EQ{\label{Sdscale}
\lim_{k\to\infty}S_D(\mathbf{k})&=\frac{1}{|\mathbf{k}|}\left(\frac{1}{2\pi\sigma^2}\right)^{(3-D)/2},
}
where for $D<3$ there is enhancement by $\sigma^{-(3-D)}$ due to confinement. 

The low-D SPGPE may be summarised in the form
\EQ{\label{Genform}
(S)i\hbar d\psi(\rr,t)={}&{\cal P}_D\Bigg\{ (1-i\gamma)\LL_D\psi(\rr,t) dt+i\hbar dW_D(\rr,t)\nonumber\\
{}&+V_D(\rr,t)\psi(\rr,t)dt-\hbar\psi(\rr,t)dU_D(\rr,t)\Bigg\},
}
with potential
\EQ{\label{Vmgen}
V_D(\rr,t)&=-\hbar \int d^D  \rr^\prime  \ve_D(\rr-\rr')\nabla^\prime\cdot \mathbf{j}_D(\rr^\prime),
}
\EQ{\label{Mgen}
\ve_D(\rr)&=\frac{\cal M}{(2\pi)^D}\int d^D \kk \; e^{i\kk \cdot \rr}S_D(\mathbf{k}),
}
and noise correlation functions
\EQ{\label{dWggen}
\langle dW_D^*(\rr,t)dW_D(\rr^\prime,t)\rangle&=\frac{2k_B T}{\hbar}\gamma\delta_D(\rr',\rr)dt,\\
\label{dWMgen}
\langle dU_D(\rr,t)dU_D(\rr^\prime,t)\rangle&=\frac{2k_B T}{\hbar}\ve_D(\rr-\rr^\prime)dt,
}
where
\EQ{
\label{deltaCD}
\delta_D(\rr,\rr^\prime)&=\sum_{n}\phi^{(D)}_n(\rr)\phi^{(D)}_n(\rr^\prime)^*,
}
is the projected delta-function of the $D$-dimensional $\rC$-field region.
We have used the shorthand $\phi_n^{(D)}(\rr)$ to represent the modes in the $D$-dimensional subspace, i.e. for a harmonic oscillator basis in $D=2$, the modes are a separable product of independent harmonic oscillator eigenstates: $\phi_n^{(2)}(\rr)\equiv \phi_{n_x}(x)\phi_{n_y}(y)$.
\subsection{Separation of length scales in 1D: White noise limit}
While the expressions \eref{1Dget}, \eref{S1def}, and \eref{dWe1} give a formally exact reduction to 1D, a further approximation can be identified in the regime of high anisotropy where the length scale $\sigma$ becomes much smaller than the length scales over which $\psi(x,t)$ changes. The scattering kernel may then be approximated as
\EQ{\label{S1hom}
S_1(k)\simeq S_1(0)=\frac{1}{\sqrt{8\pi\sigma^2}},
}
giving the potential 
\EQ{\label{V1hom}
V_1(x,t)&\simeq\bar{V}_1(x,t)=-\hbar\bar{\cal M}\partial_{x}j(x),
}
with
\EQ{\label{mbar}
\bar{\cal M}&\equiv\frac{{\cal M}}{\sqrt{8\pi\sigma^2}}.
}
Making use of the separation of scales to simplify the noise, we arrive at
\EQ{\label{M1Dhom}
\ve_1(x)&\simeq\bar{\ve}_1 (x)\equiv\bar{\cal M}\delta(x),
}
and the white noise correlation function
\EQ{\label{dWM1Dhom}
\langle dU_1(x,t)dU_1(x',t)\rangle&\approx\langle d\bar{U}_1(x,t)d\bar{U}_1(x',t)\rangle\nonumber\\
&\equiv\frac{2k_B T}{\hbar}\bar{\cal M}\delta(x-x')dt.
}
Thus, for large enough anisotropy the energy-damping noise becomes broadband white noise.

\section{Variational Analysis}\label{sec:varVal}
In this section we present an assessment of the validity criteria \eref{2Dcond}, \eref{1Dcond} for the effective low-dimensional stochastic equations of motion. 

\subsection{Estimating SPGPE parameters \label{sec:eqn_of_state}}
A basic problem in SPGPE theory involves determining parameters for modelling experiments, as has been studied in 3D~\cite{Rooney:2010dp}. In general we assume the experimental parameters take the form of a definite total particle number $N$, and temperature $T$, and we wish to find $\mu(N,T)$ and $\ec(N,T)$ suitable for modelling the system using SPGPE theory. For the low-$D$ effective theories, the value of $\ec$ is partially constrained by the geometry, requiring that we assess the validity of projecting into the low-$D$ subspace.
\par
To determine system parameters, we start from an ideal gas description of the thermal cloud in a 3D parabolic trap, with BEC transition occurring at temperature
\EQ{\label{Tc0}
k_B T_c^0 &=  \hbar \bar{\omega} \left(\frac{N}{\zeta(3)}\right)^{1/3},
} 
for a given total particle number $N$, and geometric-mean trap frequency $\bar\omega^3=\omega_x\omega_y\omega_z$. For highly anisotropic confinement we must also include the finite-size and interaction shifts in the determination of $T_c$, giving the transition temperature
\EQ{\label{TcEst}
T_{c}(N) &= T_c^0 + \delta T_c^{\rm fs} + \delta T_c^{\rm int},
}
for finite-size and interactions shifts~\cite{Dalfovo1999}
\EQ{\label{fsT}
\delta T_c^{\textrm{fs}} &= - \frac{\zeta(2)}{2\zeta(3)^{2/3}} \frac{\omega}{\bar{\omega}} N^{-1/3} T_c^0,\\
\label{intT}
\delta T_c^{\textrm{int}} &= -1.33 \frac{a}{\bar{a}} N^{1/6} T_c^0.
}
Here $\omega = (\omega_x + \omega_y + \omega_z)/3$, and  $\bar{a}^3 = a_xa_ya_z$ is the geometric mean of the oscillator lengths $a_x=\sqrt{\hbar/m\omega_x}$, etc. For a given atom number $N$, we estimate the condensate number via the ideal gas equation of state 
\EQ{\label{idealG}
N_0/N&=1-(T/T_c )^3.
}
\par
For $N_0$ atoms in a low-D Thomas-Fermi state, we use \eref{gD} to find the chemical potentials
 \EQ{\label{muTF}
\mu_D^0&=\begin{cases}
\hbar\omega_\perp\left(\sqrt{\frac{8}{\pi}}\frac{a N_0}{\sigma }\right)^{1/2}& (D=2),\\
\hbar\omega_z\left(\frac{3}{\sqrt{8}}\frac{a a_z N_0}{\sigma^2}\right)^{2/3}& (D=1).
\end{cases}
}
We then use \eref{muD} to estimate the 3D chemical potential as
\EQ{\label{mu3D}
\mu(N,T)&=\begin{cases}
\mu_2^0+\hbar\omega_z/2& (D=2),\\
\mu_1^0+\hbar\omega_\perp& (D=1).
\end{cases}
 }

We determine the $D$-dimensional energy cutoff by inverting the Bose-Einstein distribution
\EQ{\label{ecut} \epsilon_{{\rm cut}}^D = \log\left( \frac{1}{1+n^D_{\rm cut}} \right) k_BT + \mu_D,}
where the population of modes at the cutoff energy and  typically chosen to be $n^D_{\rm cut} \sim 1-3$, consistent with the truncated Wigner approximation~\cite{Blakie:2008is}. This choice of $\ec^D$ ensures that the number of modes in the weakly confined direction is preserved when the dimensional reduction is made.
 
\subsection{Corrections to the transverse ground state: hybrid Lagrangian variational method \label{sec:HVLM}}
The low-$D$ chemical potentials given by \eref{muTF} describe the regime where the transverse wavefunction is in the ground state. The regime of validity for this approximation can be understood by determining the leading order correction to this expression, for a transverse wavefunction that is allowed to vary, while remaining Gaussian. A variational treatment of dynamics with such a transverse wavefunction leads to the \emph{hybrid Lagrangian variational method} (HLVM)~\cite{Edwards:2012dh}, giving an effective Gross-Pitaevskii equation with modified interactions, coupled to an auxiliary equation of motion for the transverse width $\lambda_D$. This description provides an estimate of the importance of post ground-state corrections in 1D and 2D.

The HLVM method seeks to find a low dimensional effective GPE, starting from the 3D GPE Lagrangian density 
\EQ{\label{GPEL}
{\cal L}&\equiv \frac{i\hbar}{2}\left(\Phi\Phi^*_t-\Phi_t\Phi^*\right)+\frac{\hbar^2}{2m}|\nabla\Phi|^2+V_T(\rr)|\Phi|^2+\frac{u}{2}|\Phi|^4.
}
Applying the variational principle to \eref{GPEL}, one obtains the Euler-Lagrange equation
\EQ{\label{ELE1}
\sum_{\eta=x,y,z,t}\partial_\eta\left(\frac{\partial {\cal L}}{\partial \Phi_\eta^*}\right)=\frac{\partial {\cal L} }{\partial \Phi^*},
}
which leads to the 3D GPE as equation of motion for the field $\Phi$. The basic approach in Ref.~\cite{Edwards:2012dh}, extended here to $D=1$, is to assume a trial wavefunction of the form
\EQ{\label{Gtrans}
\Phi(\rr,t)&=\begin{cases}
\psi(x,y,t)A(t)\exp{\left(-z^2/2\lambda^2+i\kappa z^2\right)}& (D=2),\\
\psi(z,t)A(t)\exp{\left(-r_\perp^2/2\lambda^2+i\kappa r_\perp^2\right)}& (D=1),
\end{cases}
} 
describing the lowest energy (breathing) excitation in the transverse dimensions.
Integrating out the transverse dimensions, we then find the Lagrangian densities
\EQ{
{\cal L}_D={}&
\frac{i\hbar}{2}\left(\psi\psi^*_t-\psi_t\psi^*\right)+\frac{\hbar^2}{2m}|\nabla_D\psi|^2+V_D(\rr,t)|\psi|^2+\frac{g_D(t)}{2}|\psi|^4\nonumber\\
\label{LagD}
{}&+\Theta_D(t)|\psi|^2,
}
where $\nabla_D^2$ gives the Laplacian in $D$ dimensions, 
\EQ{\label{gDdef}
g_D(t)&=\frac{u}{(2\pi\lambda(t)^2)^{(3-D)/2}},
}

is the effective interaction, and
\EQ{\label{ThetaD}
\Theta_D(t)&=
\begin{cases}
\frac{\hbar\lambda^2\dot\kappa}{2}+\frac{\hbar^2}{4m\lambda^2}+\frac{\hbar^2\lambda^2\kappa^2}{m}+\frac{m\omega_z^2\lambda^2}{4}& (D=2),\\
\hbar\lambda^2\dot\kappa+\frac{\hbar^2}{2m\lambda^2}+\frac{2\hbar^2\lambda^2\kappa^2}{m}+\frac{m\omega_\perp^2\lambda^2}{2}& (D=1).
\end{cases}
}
contains all of the transverse energy terms.
The Euler-Lagrange equations are now
\EQ{\label{redELE1}
\partial_t\left(\frac{\partial {\cal L}_D}{\partial \dot\kappa}\right)&=\frac{\partial {\cal L}_D}{\partial \kappa},\\
\label{redELE2}
\partial_t\left(\frac{\partial {\cal L}_D}{\partial \dot\lambda}\right)&=\frac{\partial {\cal L}_D}{\partial \lambda},\\
\label{redELE3}
\sum_{\eta'}\partial_{\eta'}\left(\frac{\partial {\cal L}}{\partial \psi_{\eta'}^*}\right)&=\frac{\partial {\cal L} }{\partial \psi^*},
}
where $\eta'$ runs over $t$ and the spatial coordinates of the $D$-dimensional subspace.
For either dimension, using \eref{redELE1}, integrating the result $d^D\rr$, and using $\partial_tN=0$, gives $\kappa=m\dot\lambda/2\hbar\lambda$; this expression allows the simplification of the $\lambda$ equation from \eref{redELE2}, as in Ref.~\cite{Edwards:2012dh}. The end result of this procedure is the set of coupled equations of motion
\EQ{\label{psiEom}
i\hbar\frac{\partial \bar\psi}{\partial t}&=\left(-\frac{\hbar^2\nabla_D^2}{2m}+V_D(\rr,t)+g_D(t)|\bar\psi|^2\right)\bar\psi,\\
\label{sigEom}
\ddot\lambda+\Omega_D^2\lambda&=\frac{\hbar^2}{m^2\lambda^3}+\frac{g_D(t)C_D(t)}{\lambda m N},
}
where 
\EQ{\label{Cdef}
C_D(t)&=\int d^D\rr\;|\bar\psi(\rr,t)|^4,
}
and to arrive at Eq.~\eref{psiEom} we we have transformed the wavefunction to the rotating frame $\bar\psi \equiv \psi \exp{\left(-i\int_0^t dt'\Theta_D(t')\right)}$. The time-independent equations
\EQ{\label{psiTI}
\mu_D\bar\psi&=\left(-\frac{\hbar^2\nabla_D^2}{2m}+V_D(\rr)+g_D|\bar\psi|^2\right)\bar\psi,\\
\label{sigTI}
\Omega_D^2 \lambda&=\frac{\hbar^2}{m^2\lambda^3}+\frac{g_DC_D}{\lambda m N},
}
can now be solved in the Thomas-Fermi approximation for $N\equiv N_0$ atoms in a Thomas-Fermi condensate with chemical potential $\mu_D^0$. For a parabolic trap, we make use of \eref{muTF} (with $\sigma_D$ replaced with variational parameter $\lambda_D$) to obtain expressions for $N_0$, and find the Thomas-Fermi expressions
\EQ{\label{CDTF}
C_D&=\begin{cases}
\frac{(2\pi)^2}{m\omega_\perp^2}\frac{(\mu_2^0)^3\lambda^2}{u^2}& (D=2),\\
\frac{16\sqrt{2}(2\pi)^2}{15}\frac{(\mu_1^0)^{5/2}\lambda^4}{u^2\sqrt{m\omega_z^2}}& (D=1).
\end{cases}
}
We can then obtain the simple result
\EQ{\label{ratio1}
\frac{g_DC_D}{ N_0}&=\alpha_D \mu_D^0.
}
For the case of parabolic confinement, the trap and geometry-dependent numerical factor is 
\EQ{\label{lamd}
\alpha_D=\begin{cases}
2/3& (D=2),\\
4/5& (D=1).
\end{cases}
}
Using \eref{ratio1} in \eref{sigTI}, and denoting the stationary solution for $\lambda_D$ in Thomas-Fermi approximation by $\bar\lambda_D$, we find the quadratic equation $\bar\lambda_D^4-\bar\lambda_D^2\alpha_D\mu_D^0/m-\hbar^2/m^2\Omega_D^2=0$, with solution
\EQ{\label{lamSol}
\bar\lambda_D^2&=\sigma_D^2\left[1+\left(\frac{\alpha_D\mu^0_D}{2\hbar\Omega_D}\right)^2\right]^{1/2}+\sigma_D^2\frac{\alpha_D\mu^0_D}{2\hbar\Omega_D},
}
where $\Omega_D$ and $\sigma_D$ are defined in \eref{OmD}, and \eref{sigD}. Thus, to leading order
\EQ{\label{lamExp}
\bar\lambda_D^2&\simeq\sigma_D^2\left(1+\frac{\alpha_D\mu^0_D}{2\hbar\Omega_D}\right),
}
and the low-dimensional regime will be reached when the second term is much less than unity. We can thus limit the fractional change in $\sigma_D^2$ to be less than $\delta$ by imposing the restriction
\EQ{\label{muC}
\mu_D^0&\leq \frac{2\hbar\Omega_D}{\alpha_D}\delta.
}
For parabolic trapping, a choice of $\delta=0.2$ translates to the constraints $\mu_1\leq \hbar\omega_\perp/2$ and $\mu_2\leq \hbar\omega_z/5$, and as we shall see in the next section, the boundary of the 1$D$ regime is well approximated by this choice. In general the specific trapping geometry of the low-$D$ space must be considered to determine $\alpha_D$.
\section{1D Regime of Validity}\label{1dval}
In this section we test the 1D validity criteria by considering the equilibrium state of a finite temperature Bose gas in a cigar trap, as found from numerical simulation of the 1D-SPGPE, and compare the results with the HLVM criteria \eref{muC}, and with the 3D-SPGPE. 
\begin{figure}[!t]
\begin{center}
\includegraphics[width=\columnwidth]{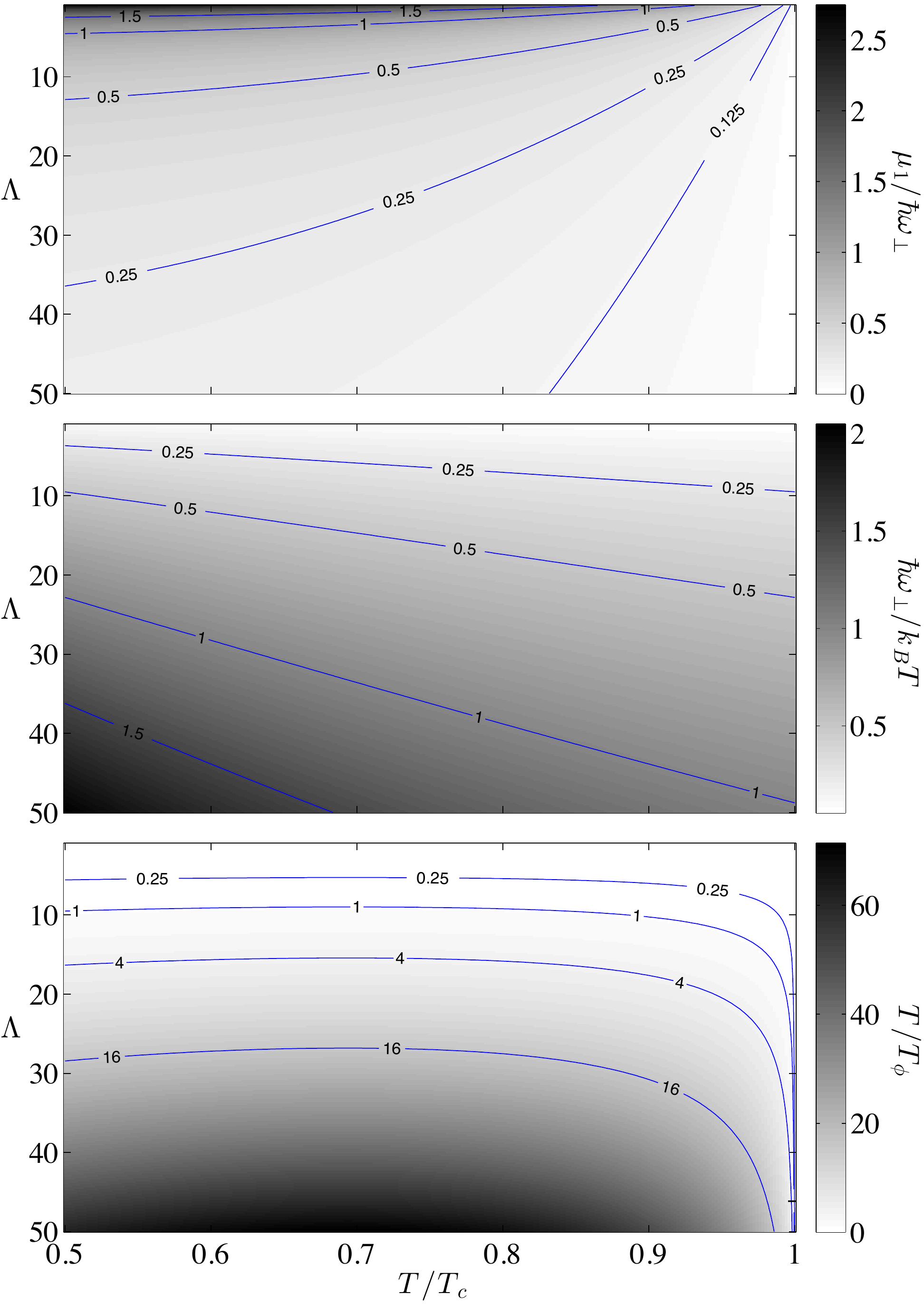}
\caption{Phase plots demonstrating the regime of validity for the one-dimensional SPGPE theory, for a range of relative temperatures and length aspect ratios ($\Lambda$). Both the total atom number ($N = 1\times10^4$) and $\omega_x = 2\pi~{\rm Hz}$ are fixed for all $T$ and $\Lambda$, and $\mu_1$ is calculated via \eref{muTF}.}
\label{fig:inequalities}
\end{center}
\end{figure}

\subsection{1D System \label{sec:num_system}}

In order to test the validity of our low-D description, we consider the dimensional reduction of a cigar-shaped trap. The trapping potential of the system is given by
\EQ{V(\rr) &= \frac{m}{2} \left( \omega_x^2 x^2 + \omega_\perp^2 (y^2 + z^2)  \right),
}
where $\omega_x \ll \omega_\perp$. We use a system consisting of $N \approx 1\times10^4$ total number of \Rb~atoms held in a trap with constant $\omega_x = 2\pi$~Hz. In terms of the $C$-region and $I$-region, this is comprised of $N=N_D+N_I$. To investigate the transition towards the 1D regime, we vary $\omega_\perp$ via the aspect ratio parameter
\EQ{\Lambda \equiv \frac{a_z}{a_\perp} = \frac{\omega_\perp^2}{\omega_z^2},}
so that as $\Lambda$ increases the system becomes increasingly prolate.
\subsection{SPGPE regimes}
In this section we highlight the different regimes which occur within the SPGPE theory during the transition to the 1D regime. We calculate the chemical potential via Eq.~\eref{muTF}, for $1 \leq \Lambda \leq 50$, $0.5 \leq T/T_c \leq 1$. 
\begin{table}[!t]
\begin{center}
\begin{tabular}{ c | c  c  c  c }
\hline
$\Lambda$ & $4$ & $8.5 $ & $14$ & $24$   \\
\hline\hline
$\mu_1/\hbar\omega_\perp$ & 0.83 & 0.50  & 0.36  & 0.25 \\
\hline
$\hbar\omega_\perp/k_B T$ & 0.18 & 0.31 & 0.45  & 0.70 \\
\hline
$\ec/\mu$ & 2.25 & 1.88 & 1.66 & 1.47 \\
\hline
$n_{\rm cut}$ [3D] &1.87  & 1.99 & 1.80 & 1.82 \\
\hline
$n_{\rm cut}$ [1D] & 1.90 & 1.97& 2.04 & 1.99  \\
\hline
$N_3 \times10^{-3}$& 7.57 & 5.93  & 5.90 & 6.05 \\
\hline
$N_1 \times10^{-3}$& 5.80 & 5.86 & 5.90 & 6.06 \\
\hline
$N_0\times10^{-3}$ [3D] & 6.22 & 2.13 & 0.75 & 0.22 \\
\hline
$N_0\times10^{-3}$ [1D] & 4.7 & 2.35 & 0.83 & 0.23 \\
\hline
$T/T_\phi$ & $0.12$ & $0.85 $ & $3.07$ & $11.95$ \\
\hline\hline
\end{tabular}
\caption{Summary of system parameters used for 1D and 3D SPGPE simulations. Equilibrium properties are also shown. For all values of $\Lambda$ we have constant $\omega_x = 2\pi~{\rm Hz}$, $N  = 1\times10^4$, and $T/T_c=0.75$. }
\label{tab:equilibrium_props}
\end{center}
\end{table}
Different regimes of validity arise from considering the 1D validity criteria inequalities in Eq.~\eref{1Dcond}. The key energy ratios are shown in \fref{fig:inequalities}. We firstly consider $\mu_1/\hbar\omega_\perp$, and see that this ratio decreases with increasing anisotropy, and most dramatically at higher temperatures.  In contrast, $\hbar\omega_\perp / k_B T$ increases with increased anisotropy, eventually suppressing the transverse thermal cloud, for large $\Lambda$. In order to make the dimensional reduction within the SPGPE framework, the system must be anisotropic enough so that the transverse wavefunction is in the ground state, but still in a regime where the thermal cloud is three-dimensional so that the SPGPE is valid. 

An important consideration is the role of phase fluctuations when approaching the quasi-1D regime. The characteristic temperature for the onset of phase fluctuations is \cite{Petrov:2000cf}
\EQ{T_{\phi} = N (\hbar \omega_x)^2 / \mu k_B \label{eq:tphi},}
which is shown in \fref{fig:inequalities}  via the ratio $T/T_\phi$; the system chemical potential $\mu$, is approximated as $\mu_1^0$ defined in \eref{muTF}. If the system is in the phase fluctuating regime, the condensate population is suppressed below the 3D ideal or interacting gas results. This does not alter the validity of the 1D-SPGPE regime, but is important to consider when interpreting the results of simulations, or in formulating approximate treatments (such as variational approaches) that require a coherent wavefunction.

\subsection{Transition to the 1D regime - SPGPE simulations \label{sec:spgpesimulations}}

We now quantitatively investigate the transition from the 3D to the 1D regime.  We calculate equilibrium states of the SPGPE for fixed $T/T_c = 0.75$, and $4 \leq \Lambda \leq 24$. Full simulation parameters are shown in Table~\ref{tab:equilibrium_props}.  We find equilibrium states of the SPGPE by evolving the simple-growth SPGPE with $\gamma = 0.4$, so that finding equilibrium only requires that we evolve the system for $t = 4$ trap cycles of evolution. 
\begin{figure}[!t]
\begin{center}
\includegraphics[width=1\columnwidth]{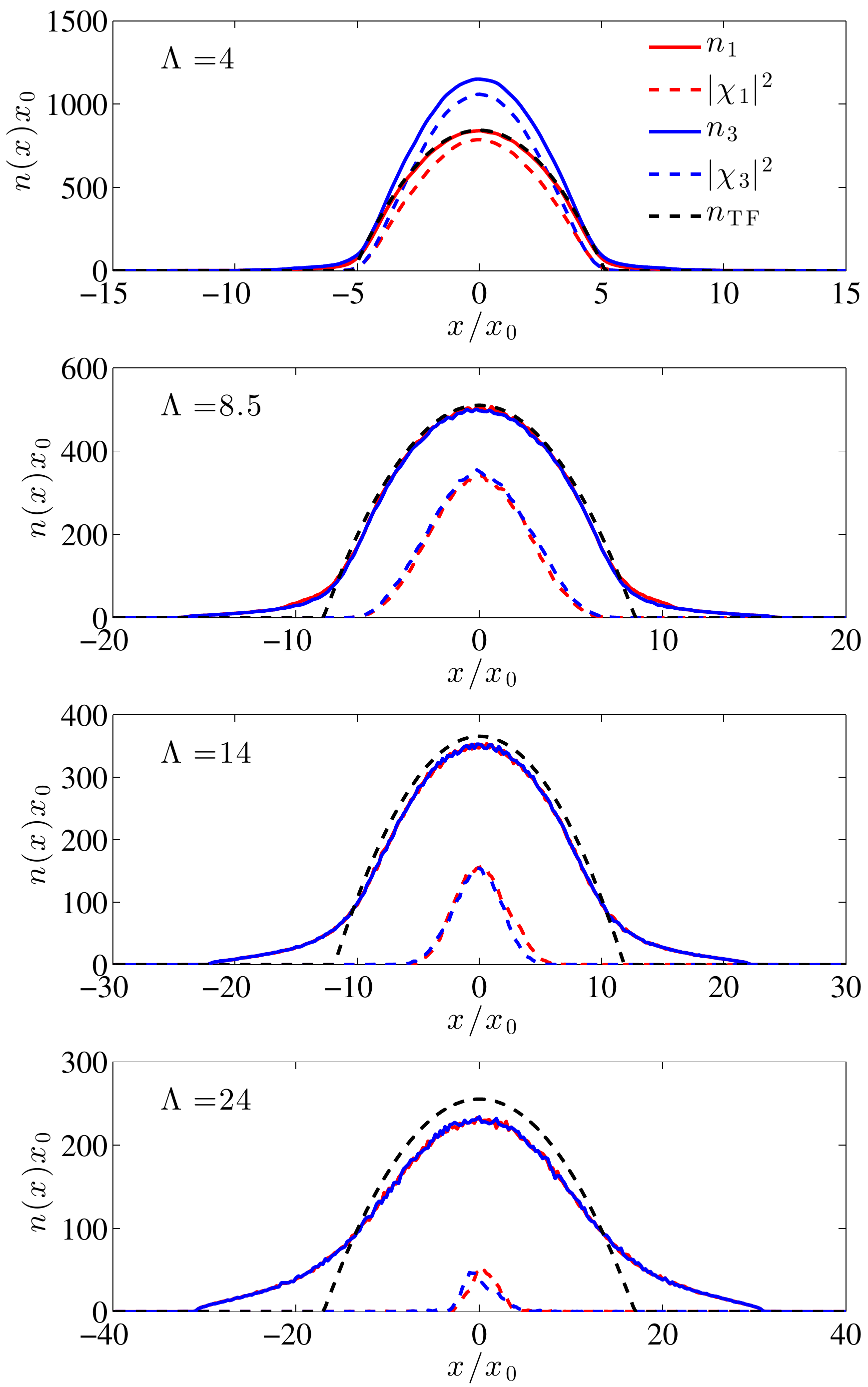}
\caption{(Color online) Ensemble averages [using 2000 trajectories] of the classical-field density profile $n_D(x)$, for a range of anisotropies at $T = 0.75T_c$, comparing the effective one-dimensional implementation (red) with three-dimensional simulations (blue).  The density corresponding to the condensate $\chi_D(x)$ is shown by the dashed red and blue lines. The 1D Thomas-Fermi density for chemical potential $\mu_1$ is shown for comparison (black dash).}
\label{fig:dens_comp_A7}
\end{center}
\end{figure}
To compare between the one-dimensional and three-dimensional equilibrium states, we consider the ensemble averaged classical-field density $n(\rr) =\langle |\psi(\rr)|^2\rangle$, and the condensate density $\chi (\rr)$  found via the Penrose-Onsager criterion:
\EQ{\label{POcrit}
\int d\rr'\rho(\rr,\rr')\chi(\rr')&=N_0\chi(\rr),
}
where the one body density matrix $\rho(\rr,\rr')\equiv\langle\psi(\rr)\psi^\dag(\rr')\rangle$ is constructed from an ensemble of 2000 trajectories for each parameter set, and the largest eigenvalue $N_0$ gives the condensate occupation. Note that the ideal gas estimate for the condensate population is $N_0=5.7\times 10^3$ for all values of $\Lambda$, and our system parameters are chosen so that finite-size and interaction shifts to $T_c$ are taken into account [see \eref{TcEst}], to ensure that $T/T_c$ stays constant (as does $N =N_D+N_I$) as $\Lambda$ increases. 

To quantitatively compare three-dimensional and one-dimensional densities, we integrate out the tightly confined dimensions of the three-dimensional density, i.e.
\EQ{n_3(x) = \int dy  \int dz \, \langle |\psi(x,y,z)|^2\rangle.} 
This projection extracts the correct 1D density from the 3D theory, for comparison with our effective low-D theory.

In \fref{fig:dens_comp_A7} we compare the total c-field and condensate densities, for the 3D and 1D simulations. For the smallest aspect ratio, $\Lambda = 4$, we find that there is a significant difference between the 3D and 1D density profiles.  The average c-field number $N_C$, and condensate number $N_0$, are shown in Table~\ref{tab:equilibrium_props}, and for this aspect ratio both the 1D-SPGPE values are $\sim24\%$ smaller than those of the 3D-SPGPE.  The large discrepancy stems from the inconsistency of the dimensional reduction: the choice of cutoff energy includes some transverse modes in the c-field, thus incorporating modes of the reservoir into the 1D-SPGPE description. Increasing the aspect ratio to $\Lambda=8.5$ | the value that ensures that the inequality \eref{muC} holds | the population artefact vanishes, and we find excellent agreement between particle densities in the 1D- and 3D-SPGPE. The condensate mode density is also closely comparable, and the c-field density agrees quite well with the 1D Thomas-Fermi result. The ratio $T/T_\phi=0.85\lesssim 1$, and the condensate population is suppressed below the ideal gas estimate, but the 1D and 3D-SPGPE values for $N_0$ are in close agreement. For $\Lambda=14$ the condensate population shows an order of magnitude departure from the 3D ideal gas result expected for a phase coherent system, consistent with $T/T_\phi=3.07\gg1$. The 1D-SPGPE describes this situation reliably, as seen in Fig.~\ref{fig:dens_comp_A7} where the 3D and 1D condensate densities agree fairly well given the finite ensemble. For $\Lambda=24$ the system is in the phase fluctuating regime $T/T_\phi\sim 12$, the condensate is strongly suppressed, and the thermal tails of the total density are a dominant feature of the distirbution. Note that for all values of $\Lambda\geq 8.5$ the 3D- and 1D-SPGPE give quite similar results for $N_C$ and $N_0$, as shown in Table \ref{tab:equilibrium_props}. We have also checked the cutoff dependence of the 1D-SPGPE, as described in the Appendix, where we find that the cutoff can be chosen consistently within the classical field approximation~\cite{Blakie:2008is} by setting the cutoff mode population $n_{\rm cut}\sim 2$.

\begin{figure}[!t]
\begin{center}
\includegraphics[width=\columnwidth]{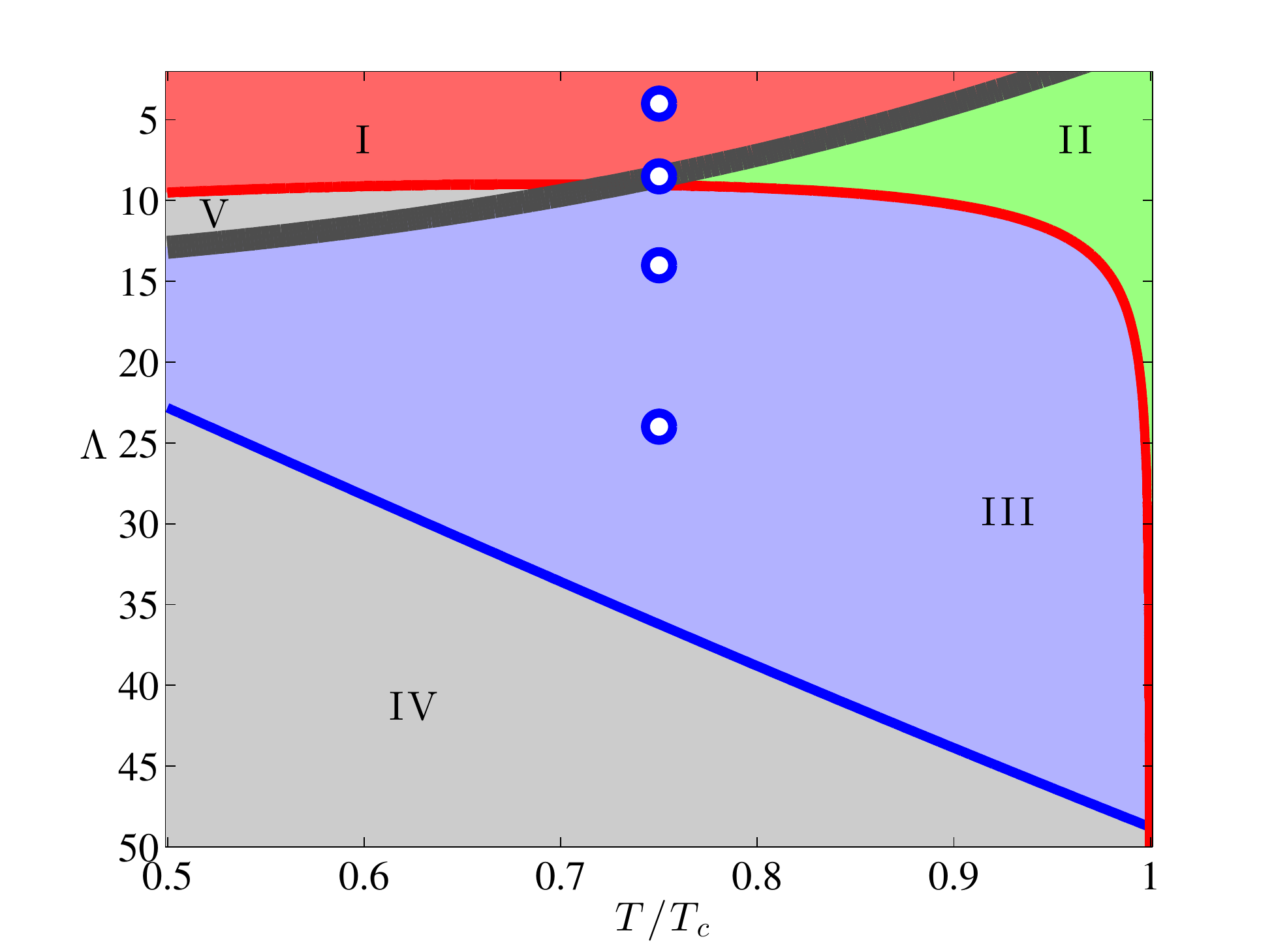}
\caption{(Color online) Summary of the different regimes for a quasi-1D system as a function of $T/T_c$ and $\Lambda$. The regimes as described in the main text are separated by the lines correspond to $T/T_\phi = 1$ (red line), $\hbar \omega_\perp/k_B T = 1$ (blue line), and $\mu_1 / \hbar \omega_\perp = 1/2$ (thick grey line). The circles correspond to SPGPE parameters used in the simulations of section \ref{sec:spgpesimulations} [See Table \ref{tab:equilibrium_props}].}
\label{fig:regimes}
\end{center}
\end{figure}
\subsection{Summary and Discussion}
The validity criteria \eref{2dsubsys} and \eref{muC}, and the results of our simulations are summarized in \fref{fig:regimes}, where we may identify several regimes:
\par 
I.~\textit{3D SPGPE.|} The system contains a 3D thermal cloud since $\hbar \omega_\perp < k_B T$. The superfluid is also in a 3D regime with $\mu_1 / \hbar \omega_\perp > 1/2$, and a well defined condensate exists since $T < T_\phi$. Thus this regime requires the 3D SPGPE for its description. 
\par
II.~\textit{Phase coherent 1D SPGPE.|} Here $\mu_1 / \hbar \omega_\perp < 1/2$ and thus the superfluid is 1D, and $\mu_1 / \hbar \omega_\perp < 1/2$, justifying the dimensional reduction. However the system remains outside the phase fluctuating regime since $T<T_\phi$.  The precise value of $\mu_1 / \hbar \omega_\perp$ for this transition is unclear, and we represent this uncertainty using a thick line in \fref{fig:regimes}. 
\par
III.~\textit{ Phase-fluctuating 1D SPGPE.|} In this regime $T > T_\phi$, and significant phase fluctuations suppress the condensate population relative to what may be expected from the 3D ideal gas equation of state \eref{idealG}. The reduction to a 1D SPGPE remains valid, as $\hbar\omega_\perp<k_BT$ so that the thermal cloud is 3D.
\par
IV.~\textit{Invalid 1D SPGPE.|} Here we have $k_B T<\hbar \omega_\perp$, and the thermal cloud loses its 3D character. The dimensionally reduced description derived in the present work is no longer valid in this regime; an SPGPE theory could be obtained  by accounting for reservoir interactions between c-field atoms and atoms of a 1D reservoir. We emphasize that the 1D-SPGPE still provides a valid means of creating equilibrium ensembles, even though it no longer describes the real-time evolution of the 1D system.
\par
V.~\textit{Phase fluctuating 3D SPGPE.|} The condensate is suppressed by phase fluctuations in this regime due to the high anisotropy and low temperature, despite the system satisfying the 3D thermal cloud criterion $\hbar\omega_\perp<k_BT$, and containing a 3D superfluid.
\begin{figure}[!t]
\begin{center}
\includegraphics[width=1\columnwidth]{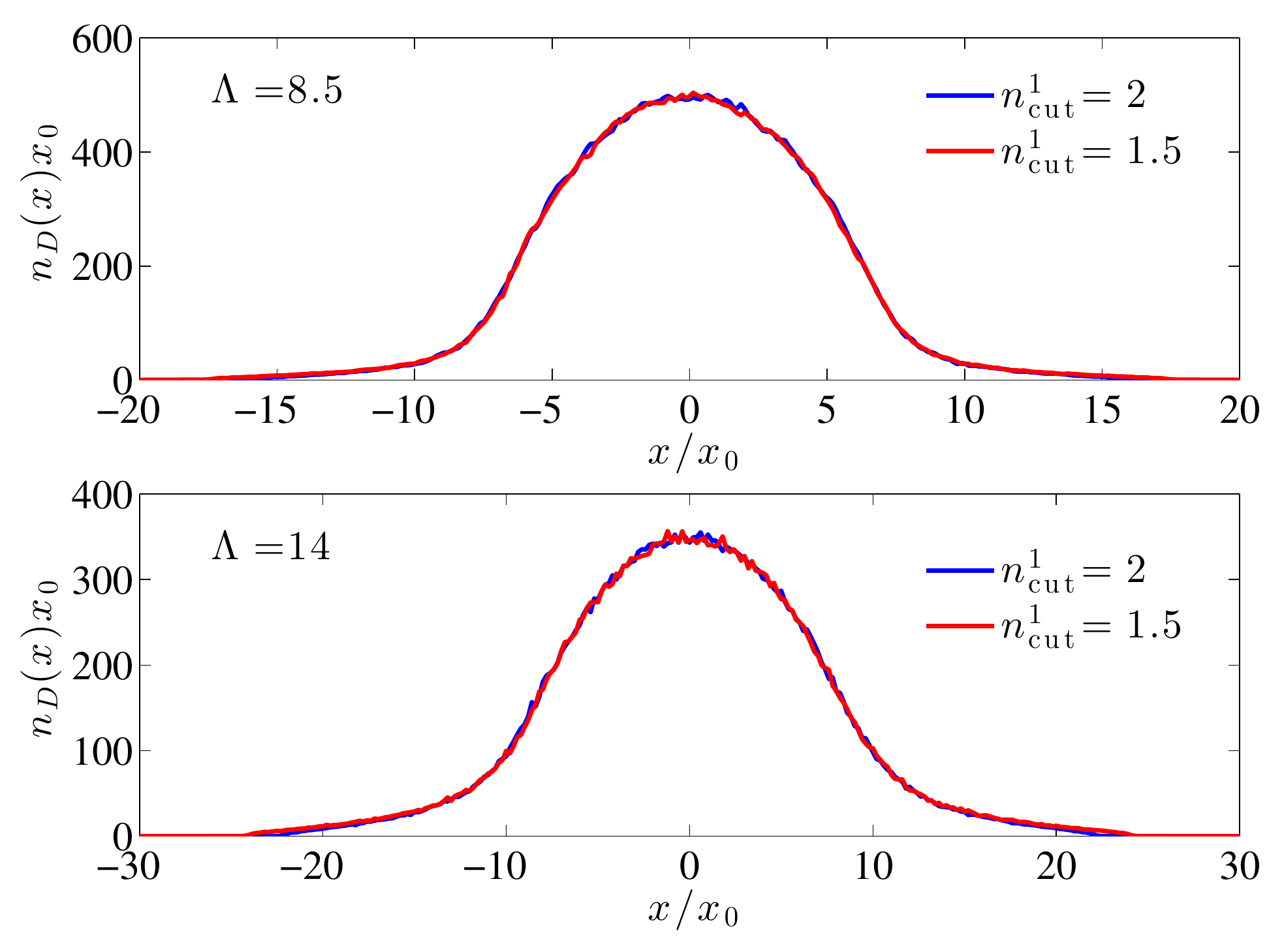}
\caption{(Color online) Average c-field density of 1D-SPGPE equilibrium states, comparing different mode populations at the cutoff energy, for $\Lambda = 8.5$ and $14$. }
\label{fig:cutoff_1D}
\end{center}
\end{figure}

\fref{fig:regimes} shows that we have a limited parameter space where the dimensional reduction is valid.  As we are always interested in a reduced dimensional c-field regime, the condition \eref{muC} is inflexible. However, the 3D thermal cloud condition $\hbar\Omega_D\lesssim k_BT$ can be relaxed for certain scenarios where an SPGPE theory still provides a valid description of a low-D subsystem.  The most obvious example is provided by a system consisting of two distinct atomic species, for which an SPGPE theory has been derived~\cite{Bradley:2014a}. For a regime of confinement (perhaps achieved through combined magneto-optical trapping) where one component is below $T_c$, and the other is above $T_c$, the number damping terms are completely suppressed, and the SPGPE takes a form where \emph{only} the energy damping reservoir interaction terms occur. The degenerate component evolves according to an SPGPE describing energy-exchange with the non-degenerate component, and the 3D thermal cloud constraint may be lifted. A related system involving buffer gas interactions as a mechanism for decoherence has been explored~\cite{Gilz:2014eo}; the master equation derived bears some resemblance to the Quantum Brownian Motion master equation governing energy damping in the two-component SPGPE theory~\cite{Bradley:2014a}.

Finally, we discuss the relationship of the SPGPE to other theoretical treatments of finite-temperature Bose-Einstein condensate (BEC) {\em dynamics}. These fall broadly into two categories: generalized mean-field theories that treat the condensate within a symmetry breaking approach~\cite{Zaremba:1999iu}, and theories stemming from a phase-space representation of the field theory~\cite{Stoof:1999tz,Gardiner:2003bk}; for reviews see~\cite{Blakie:2008is,Proukakis:2008eo}. A notable exception is given by the {\em number conserving} approach to the mean field theory, that avoids breaking $U(1)$ symmetry, while describing the dynamics of the condensate and its Bogoliubov fluctuations~\cite{Gardiner:2007gj,Billam:2013kb,Mason:2014dd}. A stochastic Gross-Pitaevskii treatment of the high temperature BEC derived via the Keldysh approach to non-equlibrium dynamics~\cite{Stoof:1999tz} has also been used in several studies~\cite{Bijlsma:2000hv,Duine:2001iu,Stoof:2001wk,Damski10a,Das:2012ki,Cockburn:2011kw,Cockburn:2012gc,Cockburn10a,Cockburn09a}. The reservoir interaction in this approach arises in a finite-temperature treatment of the many-body $T$-matrix, giving a theory of the low-energy field where fluctuations arise through the exchange of particles with the thermal fraction. This approach is essentially equivalent to any non-projected stochastic Gross-Pitaevskii equation for which the equilibrium ensemble is grand canonical~\cite{Su:2012jp,Su:2013dh}. It typically gives a very good description of equilibrium properties, and in many situations describes the same physics as the simple growth SPGPE. However, the theory lacks a formal separation of the system via an orthogonal projector. Consequently, there are challenges in handling the inherent UV-divergence in a physically and numerically consistent manner (with sufficient care this can be achieved in equilibrium \cite{Cockburn:2012gc}), and the dynamical predictions of the theory differ as the energy-damping terms are absent. 
\begin{figure}[!t]
\begin{center}
\includegraphics[width=1\columnwidth]{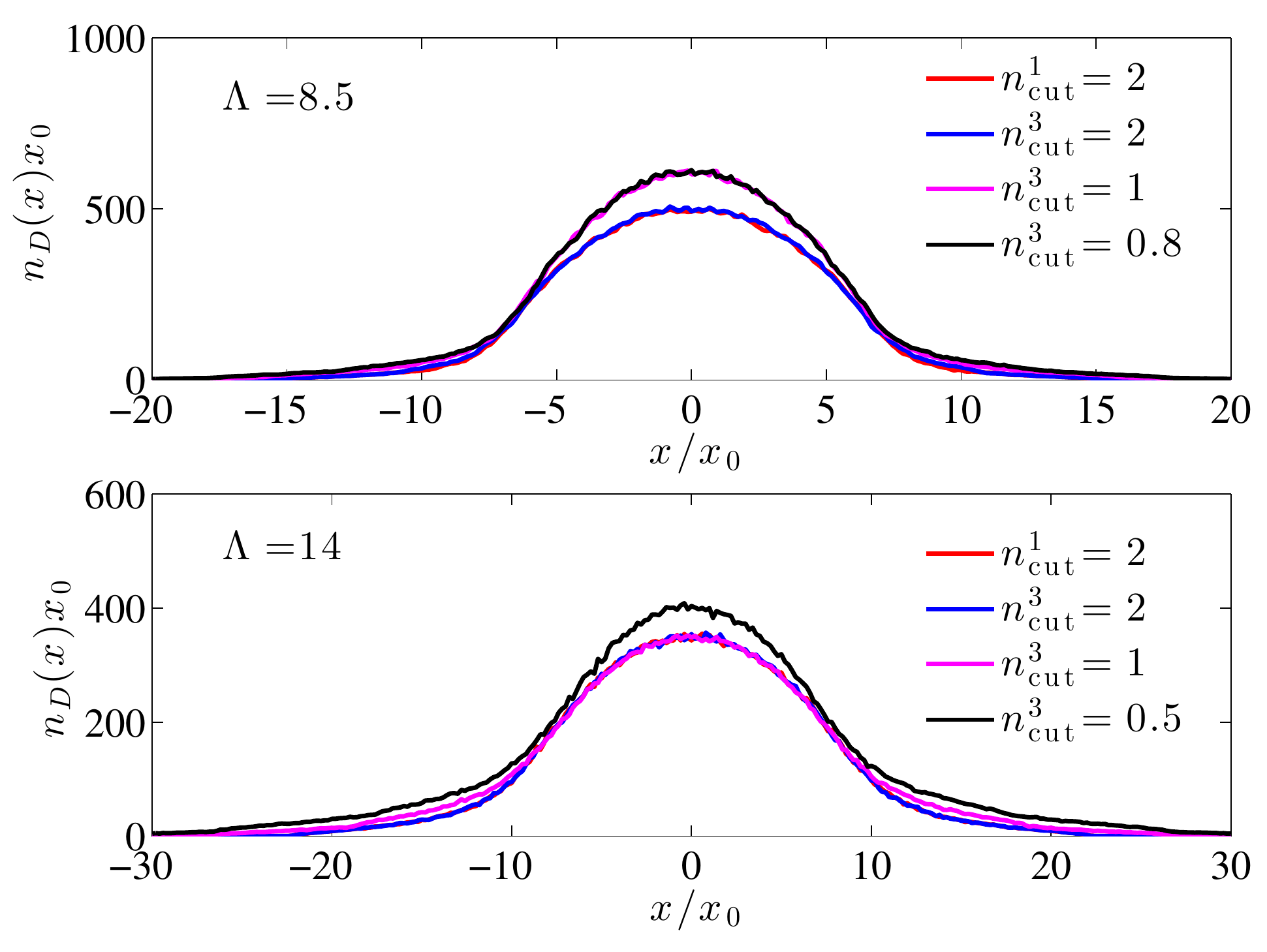}

\caption{(Color online) Average c-field density of 3D-SPGPE equilibrium states, comparing different mode populations at the cutoff energy, for $\Lambda = 8.5$ and $14$.  The 1D-SPGPE equilibrium solution is shown for comparison (red line), and in close agreement with the 3D-SPGPE for $n_{\rm cut}^3=2$ (blue line). }
\label{fig:dens1D_cutoff_8_14}
\end{center}
\end{figure}
\section{Conclusions}
The 3D-SPGPE has had some success in describing experiments with either minimal~\cite{Weiler:2008eu} or no use of fitted parameters~\cite{Rooney:2013ff}. However, the 3D theory remains numerically challenging, and is also an inconvenient framework for developing analytical models of the dissipative dynamics of excitations in finite-temperature BECs.
In this work we have presented a dimensionally reduced stochastic projected Gross-Pitaevskii equation that provides a more tractable approach for scenarios where a quantum-degenerate fraction of the atoms resides in a physical subspace of reduced spatial dimension, and the tightly confined transverse dimensions are well described by the quantum mechanical ground state of the system (assumed to be Gaussian). In this regime the SPGPE projection formalism for separating the system into coherent ($\rC$-field) and incoherent ($\rI$-field) regions provides a physically natural and formally rigorous way to find the low-D SPGPE.
\par
We have gone some way towards validating this approach for the 1D-SPGPE, via a variational treatment of the transverse degrees of freedom and numerical simulations comparing 3D-SPGPE with 1D-SPGPE. We find that the obvious inequality governing the regime of validity, namely that the system chemical potential, $\mu$, and transverse oscillator energy, $\hbar\omega_\perp$, should satisfy $\mu\ll \hbar\omega_\perp\ll k_BT$, is overly restrictive and can be significantly relaxed. In practice the condition for the 1D-SPGPE can be stated as $\kappa \mu\leq \hbar\omega_\perp\lesssim k_BT$, where $\kappa>1$, and the precise value is system dependent; for the cigar trap, we find empirically that $\kappa=2$ provides a reasonable guide for validity of dimensional reduction [see  \eref{muC}].

Areas of future interest include quantifying the 2D-SPGPE validity regime, application of the low-D SPGPE to the Kibble-Zurek mechanism, to the dissipative dynamics of solitons and vortices, and to two-dimensional quantum turbulence. 

\acknowledgements
We are grateful to Blair Blakie, Niels Kj\ae rgaard, and Crispin Gardiner for useful discussions.
ASB is supported by a Rutherford Discovery Fellowship.

\appendix*\label{app1}
\section{Cutoff dependence}
We have carried out two separate tests of the cutoff dependence of our results. First, we varied the specific choice of population at the cutoff energy, for the 1D-SPGPE. The results are shown in \fref{fig:cutoff_1D}, where we see essentially negligible change in the particle density, despite changing the cutoff population by 33\%. Second, we performed 3D-SPGPE simulations for different choices of cutoff population and compared with the 1D-SPGPE results, revealing a significant cutoff dependence in the  3D-SPGPE results, as shown in \fref{fig:dens1D_cutoff_8_14}. This sensitivity is essentially the same mechanism behind the jump in \fref{fig:dens_comp_A7} for $\Lambda=4$: increasing the cutoff energy above $ 2\hbar\omega_\perp$ will pick up the first transverse excited state in the c-field description, and so some care is required in selecting a 3D energy cutoff that generates a consistent projection into the transverse ground state. For $\Lambda=8.5$ a choice of cutoff mode population $n_{\rm cut}^3=0.8$ or $n_{\rm cut}^3=1$ selects an energy higher than required for the projection onto the transverse ground state, while  $n_{\rm cut}^3=2$ gives a c-field density in close agreement with the 1D-SPGPE. For $\Lambda=14$ the behaviour is very similar, but now convergence to the 1D-SPGPE regime is already apparent at $n_{\rm cut}^3=1$; true convergence in the tails of the distribution is seen for $n_{\rm cut}^3=2$.


\begin{thebibliography}{10}%
\makeatletter
\providecommand \@ifxundefined [1]{%
 \ifx #1\undefined \expandafter \@firstoftwo
 \else \expandafter \@secondoftwo
\fi
}%
\providecommand \@ifnum [1]{%
 \ifnum #1\expandafter \@firstoftwo
 \else \expandafter \@secondoftwo
\fi
}%
\providecommand \enquote [1]{``#1''}%
\providecommand \bibnamefont  [1]{#1}%
\providecommand \bibfnamefont [1]{#1}%
\providecommand \citenamefont [1]{#1}%
\providecommand\href[0]{\@sanitize\@href}%
\providecommand\@href[1]{\endgroup\@@startlink{#1}\endgroup\@@href}%
\providecommand\@@href[1]{#1\@@endlink}%
\providecommand \@sanitize [0]{\begingroup\catcode`\&12\catcode`\#12\relax}%
\@ifxundefined \pdfoutput {\@firstoftwo}{%
 \@ifnum{\z@=\pdfoutput}{\@firstoftwo}{\@secondoftwo}%
}{%
 \providecommand\@@startlink[1]{\leavevmode}%
 \providecommand\@@endlink[0]{}%
}{%
 \providecommand\@@startlink[1]{%
  \leavevmode
  \pdfstartlink
   attr{/Border[0 0 1 ]/H/I/C[0 1 1]}%
   user{/Subtype/Link/A<</Type/Action/S/URI/URI(#1)>>}%
  \relax
 }%
 \providecommand\@@endlink[0]{\pdfendlink}%
}%
\providecommand \url  [0]{\begingroup\@sanitize \@url }%
\providecommand \@url [1]{\endgroup\@href {#1}{\urlprefix}}%
\providecommand \urlprefix [0]{URL }%
\providecommand \Eprint[0]{\href }%
\@ifxundefined \urlstyle {%
  \providecommand \doi [1]{doi:\discretionary{}{}{}#1}%
}{%
  \providecommand \doi [0]{doi:\discretionary{}{}{}\begingroup
  \urlstyle{rm}\Url }%
}%
\providecommand \doibase [0]{http://dx.doi.org/}%
\providecommand \Doi[1]{\href{\doibase#1}}%
\providecommand \bibAnnote [3]{%
  \BibitemShut{#1}%
  \begin{quotation}\noindent
    \textsc{Key:}\ #2\\\textsc{Annotation:}\ #3%
  \end{quotation}%
}%
\providecommand \bibAnnoteFile [2]{%
  \IfFileExists{#2}{\bibAnnote {#1} {#2} {\input{#2}}}{}%
}%
\providecommand \typeout [0]{\immediate \write \m@ne }%
\providecommand \selectlanguage [0]{\@gobble}%
\providecommand \bibinfo [0]{\@secondoftwo}%
\providecommand \bibfield [0]{\@secondoftwo}%
\providecommand \translation [1]{[#1]}%
\providecommand \BibitemOpen[0]{}%
\providecommand \bibitemStop [0]{}%
\providecommand \bibitemNoStop [0]{.\EOS\space}%
\providecommand \EOS [0]{\spacefactor3000\relax}%
\providecommand \BibitemShut [1]{\csname bibitem#1\endcsname}%
\bibitem{Gardiner:2003bk}%
  \BibitemOpen
  \bibfield{author}{%
  \bibinfo {author} {\bibfnamefont{C.~W.}\ \bibnamefont{Gardiner}}\ and\
  \bibinfo {author} {\bibfnamefont{M.~J.}\ \bibnamefont{Davis}},\ }%
  \bibfield{journal}{%
  \Doi{10.1088/0953-4075/36/23/010}{\bibinfo {journal} {J. Phys. B: At. Mol.
  Opt. Phys.}}\ }%
  \textbf{\bibinfo {volume} {36}},\ \bibinfo {pages} {4731} (\bibinfo {year}
  {2003}),\ \url{http://iopscience.iop.org/0953-4075/36/23/010}%
  \bibAnnoteFile{NoStop}{Gardiner:2003bk}%
\bibitem{QKI}%
  \BibitemOpen
  \bibfield{author}{%
  \bibinfo {author} {\bibfnamefont{C.~W.}\ \bibnamefont{Gardiner}}\ and\
  \bibinfo {author} {\bibfnamefont{P.}~\bibnamefont{Zoller}},\ }%
  \bibfield{journal}{%
  \Doi{10.1103/PhysRevA.55.2902}{\bibinfo {journal} {Phys. Rev. A}}\ }%
  \textbf{\bibinfo {volume} {55}},\ \bibinfo {pages} {2902} (\bibinfo {year}
  {1997}),\ \url{http://link.aps.org/doi/10.1103/PhysRevA.55.2902}%
  \bibAnnoteFile{NoStop}{QKI}%
\bibitem{QKIII}%
  \BibitemOpen
  \bibfield{author}{%
  \bibinfo {author} {\bibfnamefont{C.~W.}\ \bibnamefont{Gardiner}}\ and\
  \bibinfo {author} {\bibfnamefont{P.}~\bibnamefont{Zoller}},\ }%
  \bibfield{journal}{%
  \Doi{10.1103/PhysRevA.58.536}{\bibinfo {journal} {Phys. Rev. A}}\ }%
  \textbf{\bibinfo {volume} {58}},\ \bibinfo {pages} {536} (\bibinfo {month}
  {Jul.}\ \bibinfo {year} {1998}),\
  \url{http://link.aps.org/doi/10.1103/PhysRevA.58.536}%
  \bibAnnoteFile{NoStop}{QKIII}%
\bibitem{Gardiner:2000fi}%
  \BibitemOpen
  \bibfield{author}{%
  \bibinfo {author} {\bibfnamefont{C.~W.}\ \bibnamefont{Gardiner}}\ and\
  \bibinfo {author} {\bibfnamefont{P.}~\bibnamefont{Zoller}},\ }%
  \bibfield{journal}{%
  \Doi{10.1103/PhysRevA.61.033601}{\bibinfo {journal} {Phys. Rev. A}}\ }%
  \textbf{\bibinfo {volume} {61}},\ \bibinfo {pages} {033601} (\bibinfo {month}
  {Feb.}\ \bibinfo {year} {2000}),\
  \url{http://link.aps.org/doi/10.1103/PhysRevA.61.033601}%
  \bibAnnoteFile{NoStop}{Gardiner:2000fi}%
\bibitem{Davis2001b}%
  \BibitemOpen
  \bibfield{author}{%
  \bibinfo {author} {\bibfnamefont{M.~J.}\ \bibnamefont{Davis}}, \bibinfo
  {author} {\bibfnamefont{S.~A.}\ \bibnamefont{Morgan}},\ and\ \bibinfo
  {author} {\bibfnamefont{K.}~\bibnamefont{Burnett}},\ }%
  \bibfield{journal}{%
  \Doi{10.1103/PhysRevLett.87.160402}{\bibinfo {journal} {Phys. Rev. Lett.}}\
  }%
  \textbf{\bibinfo {volume} {87}},\ \bibinfo {pages} {160402} (\bibinfo {month}
  {Sep.}\ \bibinfo {year} {2001}),\
  \url{http://link.aps.org/doi/10.1103/PhysRevLett.87.160402}%
  \bibAnnoteFile{NoStop}{Davis2001b}%
\bibitem{Davis2002}%
  \BibitemOpen
  \bibfield{author}{%
  \bibinfo {author} {\bibfnamefont{M.~J.}\ \bibnamefont{Davis}}, \bibinfo
  {author} {\bibfnamefont{S.~A.}\ \bibnamefont{Morgan}},\ and\ \bibinfo
  {author} {\bibfnamefont{K.}~\bibnamefont{Burnett}},\ }%
  \bibfield{journal}{%
  \Doi{10.1103/PhysRevA.66.053618}{\bibinfo {journal} {Phys. Rev. A}}\ }%
  \textbf{\bibinfo {volume} {66}},\ \bibinfo {pages} {053618} (\bibinfo {month}
  {Nov.}\ \bibinfo {year} {2002}),\
  \url{http://link.aps.org/doi/10.1103/PhysRevA.66.053618}%
  \bibAnnoteFile{NoStop}{Davis2002}%
\bibitem{Blakie05a}%
  \BibitemOpen
  \bibfield{author}{%
  \bibinfo {author} {\bibfnamefont{P.~B.}\ \bibnamefont{Blakie}}\ and\ \bibinfo
  {author} {\bibfnamefont{M.~J.}\ \bibnamefont{Davis}},\ }%
  \bibfield{journal}{%
  \Doi{10.1103/PhysRevA.72.063608}{\bibinfo {journal} {Phys. Rev. A}}\ }%
  \textbf{\bibinfo {volume} {72}},\ \bibinfo {pages} {063608} (\bibinfo {year}
  {2005}),\ \url{http://prola.aps.org/abstract/PRA/v72/i6/e063608}%
  \bibAnnoteFile{NoStop}{Blakie05a}%
\bibitem{QKPRLII}%
  \BibitemOpen
  \bibfield{author}{%
  \bibinfo {author} {\bibfnamefont{C.~W.}\ \bibnamefont{Gardiner}}, \bibinfo
  {author} {\bibfnamefont{M.~D.}\ \bibnamefont{Lee}}, \bibinfo {author}
  {\bibfnamefont{R.~J.}\ \bibnamefont{Ballagh}}, \bibinfo {author}
  {\bibfnamefont{M.~J.}\ \bibnamefont{Davis}},\ and\ \bibinfo {author}
  {\bibfnamefont{P.}~\bibnamefont{Zoller}},\ }%
  \bibfield{journal}{%
  \Doi{10.1103/PhysRevLett.81.5266}{\bibinfo {journal} {Phys. Rev. Lett.}}\ }%
  \textbf{\bibinfo {volume} {81}},\ \bibinfo {pages} {5266} (\bibinfo {month}
  {Dec.}\ \bibinfo {year} {1998}),\
  \url{http://link.aps.org/doi/10.1103/PhysRevLett.81.5266}%
  \bibAnnoteFile{NoStop}{QKPRLII}%
\bibitem{QKPRLIII}%
  \BibitemOpen
  \bibfield{author}{%
  \bibinfo {author} {\bibfnamefont{M.}~\bibnamefont{Kohl}}, \bibinfo {author}
  {\bibfnamefont{M.~J.}\ \bibnamefont{Davis}}, \bibinfo {author}
  {\bibfnamefont{C.~W.}\ \bibnamefont{Gardiner}}, \bibinfo {author}
  {\bibfnamefont{T.~W.}\ \bibnamefont{H{\"a}nsch}},\ and\ \bibinfo {author}
  {\bibfnamefont{T.}~\bibnamefont{Esslinger}},\ }%
  \bibfield{journal}{%
  \Doi{10.1103/PhysRevLett.88.080402}{\bibinfo {journal} {Phys. Rev. Lett.}}\
  }%
  \textbf{\bibinfo {volume} {88}},\ \bibinfo {pages} {080402} (\bibinfo {month}
  {Feb.}\ \bibinfo {year} {2002}),\
  \url{http://link.aps.org/doi/10.1103/PhysRevLett.88.080402}%
  \bibAnnoteFile{NoStop}{QKPRLIII}%
\bibitem{Davis:2006ic}%
  \BibitemOpen
  \bibfield{author}{%
  \bibinfo {author} {\bibfnamefont{M.~J.}\ \bibnamefont{Davis}}\ and\ \bibinfo
  {author} {\bibfnamefont{P.~B.}\ \bibnamefont{Blakie}},\ }%
  \bibfield{journal}{%
  \Doi{10.1103/PhysRevLett.96.060404}{\bibinfo {journal} {Phys. Rev. Lett.}}\
  }%
  \textbf{\bibinfo {volume} {96}},\ \bibinfo {pages} {060404} (\bibinfo {month}
  {Feb.}\ \bibinfo {year} {2006}),\
  \url{http://link.aps.org/doi/10.1103/PhysRevLett.96.060404}%
  \bibAnnoteFile{NoStop}{Davis:2006ic}%
\bibitem{Blakie2007a}%
  \BibitemOpen
  \bibfield{author}{%
  \bibinfo {author} {\bibfnamefont{P.~B.}\ \bibnamefont{Blakie}}\ and\ \bibinfo
  {author} {\bibfnamefont{M.~J.}\ \bibnamefont{Davis}},\ }%
  \bibfield{journal}{%
  \Doi{10.1088/0953-4075/40/11/007}{\bibinfo {journal} {J. Phys. B: At. Mol.
  Opt. Phys.}}\ }%
  \textbf{\bibinfo {volume} {40}},\ \bibinfo {pages} {2043} (\bibinfo {month}
  {May}\ \bibinfo {year} {2007})%
  \bibAnnoteFile{NoStop}{Blakie2007a}%
\bibitem{Bezett09b}%
  \BibitemOpen
  \bibfield{author}{%
  \bibinfo {author} {\bibfnamefont{A.}~\bibnamefont{Bezett}}\ and\ \bibinfo
  {author} {\bibfnamefont{P.~B.}\ \bibnamefont{Blakie}},\ }%
  \bibfield{journal}{%
  \Doi{10.1103/PhysRevA.79.033611}{\bibinfo {journal} {Phys. Rev. A}}\ }%
  \textbf{\bibinfo {volume} {79}},\ \bibinfo {pages} {033611} (\bibinfo {year}
  {2009}),\
  \url{http://scitation.aip.org/getabs/servlet/GetabsServlet?prog=normal&id=PLRAAN000079000003033611000001&idtype=cvips&gifs=yes}%
  \bibAnnoteFile{NoStop}{Bezett09b}%
\bibitem{Bezett09a}%
  \BibitemOpen
  \bibfield{author}{%
  \bibinfo {author} {\bibfnamefont{A.}~\bibnamefont{Bezett}}\ and\ \bibinfo
  {author} {\bibfnamefont{P.~B.}\ \bibnamefont{Blakie}},\ }%
  \bibfield{journal}{%
  \Doi{10.1103/PhysRevA.79.023602}{\bibinfo {journal} {Phys. Rev. A}}\ }%
  \textbf{\bibinfo {volume} {79}},\ \bibinfo {pages} {023602} (\bibinfo {year}
  {2009}),\
  \url{http://scitation.aip.org/getabs/servlet/GetabsServlet?prog=normal&id=PLRAAN000079000002023602000001&idtype=cvips&gifs=yes}%
  \bibAnnoteFile{NoStop}{Bezett09a}%
\bibitem{Bradley:2008gq}%
  \BibitemOpen
  \bibfield{author}{%
  \bibinfo {author} {\bibfnamefont{A.~S.}\ \bibnamefont{Bradley}}, \bibinfo
  {author} {\bibfnamefont{C.~W.}\ \bibnamefont{Gardiner}},\ and\ \bibinfo
  {author} {\bibfnamefont{M.~J.}\ \bibnamefont{Davis}},\ }%
  \bibfield{journal}{%
  \Doi{10.1103/PhysRevA.77.033616}{\bibinfo {journal} {Phys. Rev. A}}\ }%
  \textbf{\bibinfo {volume} {77}},\ \bibinfo {pages} {033616} (\bibinfo {month}
  {Mar.}\ \bibinfo {year} {2008}),\
  \url{http://link.aps.org/doi/10.1103/PhysRevA.77.033616}%
  \bibAnnoteFile{NoStop}{Bradley:2008gq}%
\bibitem{Rooney:2010dp}%
  \BibitemOpen
  \bibfield{author}{%
  \bibinfo {author} {\bibfnamefont{S.~J.}\ \bibnamefont{Rooney}}, \bibinfo
  {author} {\bibfnamefont{A.~S.}\ \bibnamefont{Bradley}},\ and\ \bibinfo
  {author} {\bibfnamefont{P.~B.}\ \bibnamefont{Blakie}},\ }%
  \bibfield{journal}{%
  \Doi{10.1103/PhysRevA.81.023630}{\bibinfo {journal} {Phys. Rev. A}}\ }%
  \textbf{\bibinfo {volume} {81}},\ \bibinfo {pages} {023630} (\bibinfo {month}
  {Feb.}\ \bibinfo {year} {2010}),\
  \url{http://link.aps.org/doi/10.1103/PhysRevA.81.023630}%
  \bibAnnoteFile{NoStop}{Rooney:2010dp}%
\bibitem{Rooney:2011fm}%
  \BibitemOpen
  \bibfield{author}{%
  \bibinfo {author} {\bibfnamefont{S.~J.}\ \bibnamefont{Rooney}}, \bibinfo
  {author} {\bibfnamefont{P.~B.}\ \bibnamefont{Blakie}}, \bibinfo {author}
  {\bibfnamefont{B.~P.}\ \bibnamefont{Anderson}},\ and\ \bibinfo {author}
  {\bibfnamefont{A.~S.}\ \bibnamefont{Bradley}},\ }%
  \bibfield{journal}{%
  \Doi{10.1103/PhysRevA.84.023637}{\bibinfo {journal} {Phys. Rev. A}}\ }%
  \textbf{\bibinfo {volume} {84}},\ \bibinfo {pages} {023637} (\bibinfo {month}
  {Aug.}\ \bibinfo {year} {2011}),\
  \url{http://link.aps.org/doi/10.1103/PhysRevA.84.023637}%
  \bibAnnoteFile{NoStop}{Rooney:2011fm}%
\bibitem{QN}%
  \BibitemOpen
  \bibfield{author}{%
  \bibinfo {author} {\bibfnamefont{C.~W.}\ \bibnamefont{Gardiner}}\ and\
  \bibinfo {author} {\bibfnamefont{P.}~\bibnamefont{Zoller}},\ }%
  \emph{\bibinfo {title} {{Quantum Noise}}},\ \bibinfo {edition} {3rd}\ ed.\
  (\bibinfo {publisher} {Springer-Verlag},\ \bibinfo {address} {Berlin
  Heidelberg},\ \bibinfo {year} {2004})%
  \bibAnnoteFile{NoStop}{QN}%
\bibitem{QO}%
  \BibitemOpen
  \bibfield{author}{%
  \bibinfo {author} {\bibfnamefont{D.~F.}\ \bibnamefont{Walls}}\ and\ \bibinfo
  {author} {\bibfnamefont{G.~J.}\ \bibnamefont{Milburn}},\ }%
  \emph{\bibinfo {title} {{Quantum Optics}}},\ \bibinfo {edition} {1st}\ ed.\
  (\bibinfo {publisher} {Springer-Verlag},\ \bibinfo {address} {Berlin
  Heidelberg},\ \bibinfo {year} {1994})%
  \bibAnnoteFile{NoStop}{QO}%
\bibitem{Rooney:2012gb}%
  \BibitemOpen
  \bibfield{author}{%
  \bibinfo {author} {\bibfnamefont{S.~J.}\ \bibnamefont{Rooney}}, \bibinfo
  {author} {\bibfnamefont{P.~B.}\ \bibnamefont{Blakie}},\ and\ \bibinfo
  {author} {\bibfnamefont{A.~S.}\ \bibnamefont{Bradley}},\ }%
  \bibfield{journal}{%
  \Doi{10.1103/PhysRevA.86.053634}{\bibinfo {journal} {Phys. Rev. A}}\ }%
  \textbf{\bibinfo {volume} {86}},\ \bibinfo {pages} {053634} (\bibinfo {month}
  {Nov.}\ \bibinfo {year} {2012}),\
  \url{http://link.aps.org/doi/10.1103/PhysRevA.86.053634}%
  \bibAnnoteFile{NoStop}{Rooney:2012gb}%
\bibitem{Weiler:2008eu}%
  \BibitemOpen
  \bibfield{author}{%
  \bibinfo {author} {\bibfnamefont{C.~N.}\ \bibnamefont{Weiler}}, \bibinfo
  {author} {\bibfnamefont{T.~W.}\ \bibnamefont{Neely}}, \bibinfo {author}
  {\bibfnamefont{D.~R.}\ \bibnamefont{Scherer}}, \bibinfo {author}
  {\bibfnamefont{A.~S.}\ \bibnamefont{Bradley}}, \bibinfo {author}
  {\bibfnamefont{M.~J.}\ \bibnamefont{Davis}},\ and\ \bibinfo {author}
  {\bibfnamefont{B.~P.}\ \bibnamefont{Anderson}},\ }%
  \bibfield{journal}{%
  \Doi{10.1038/nature07334}{\bibinfo {journal} {Nature}}\ }%
  \textbf{\bibinfo {volume} {455}},\ \bibinfo {pages} {948} (\bibinfo {month}
  {Oct.}\ \bibinfo {year} {2008}),\
  \url{http://www.nature.com/doifinder/10.1038/nature07334}%
  \bibAnnoteFile{NoStop}{Weiler:2008eu}%
\bibitem{Rooney:2013ff}%
  \BibitemOpen
  \bibfield{author}{%
  \bibinfo {author} {\bibfnamefont{S.~J.}\ \bibnamefont{Rooney}}, \bibinfo
  {author} {\bibfnamefont{T.~W.}\ \bibnamefont{Neely}}, \bibinfo {author}
  {\bibfnamefont{B.~P.}\ \bibnamefont{Anderson}},\ and\ \bibinfo {author}
  {\bibfnamefont{A.~S.}\ \bibnamefont{Bradley}},\ }%
  \bibfield{journal}{%
  \Doi{10.1103/PhysRevA.88.063620}{\bibinfo {journal} {Phys. Rev. A}}\ }%
  \textbf{\bibinfo {volume} {88}},\ \bibinfo {pages} {063620} (\bibinfo {month}
  {Dec.}\ \bibinfo {year} {2013}),\
  \url{http://link.aps.org/doi/10.1103/PhysRevA.88.063620}%
  \bibAnnoteFile{NoStop}{Rooney:2013ff}%
\bibitem{Garrett:2013gk}%
  \BibitemOpen
  \bibfield{author}{%
  \bibinfo {author} {\bibfnamefont{M.~C.}\ \bibnamefont{Garrett}}, \bibinfo
  {author} {\bibfnamefont{T.~M.}\ \bibnamefont{Wright}},\ and\ \bibinfo
  {author} {\bibfnamefont{M.~J.}\ \bibnamefont{Davis}},\ }%
  \bibfield{journal}{%
  \Doi{10.1103/PhysRevA.87.063611}{\bibinfo {journal} {Phys. Rev. A}}\ }%
  \textbf{\bibinfo {volume} {87}},\ \bibinfo {pages} {063611} (\bibinfo {month}
  {Jun.}\ \bibinfo {year} {2013}),\
  \url{http://link.aps.org/doi/10.1103/PhysRevA.87.063611}%
  \bibAnnoteFile{NoStop}{Garrett:2013gk}%
\bibitem{Lee:2015vd}%
  \BibitemOpen
  \bibfield{author}{%
  \bibinfo {author} {\bibfnamefont{G.~M.}\ \bibnamefont{Lee}}, \bibinfo
  {author} {\bibfnamefont{S.~A.}\ \bibnamefont{Haine}}, \bibinfo {author}
  {\bibfnamefont{A.~S.}\ \bibnamefont{Bradley}},\ and\ \bibinfo {author}
  {\bibfnamefont{M.~J.}\ \bibnamefont{Davis}}}%
   (\bibinfo {month} {Jun.}\ \bibinfo {year} {2015}),\
  \Eprint{http://arxiv.org/abs/1506.00321}{1506.00321},\
  \url{http://arxiv.org/abs/1506.00321}%
  \bibAnnoteFile{NoStop}{Lee:2015vd}%
\bibitem{Davis:2012hq}%
  \BibitemOpen
  \bibfield{author}{%
  \bibinfo {author} {\bibfnamefont{M.~J.}\ \bibnamefont{Davis}}, \bibinfo
  {author} {\bibfnamefont{P.~B.}\ \bibnamefont{Blakie}}, \bibinfo {author}
  {\bibfnamefont{A.~H.}\ \bibnamefont{van Amerongen}}, \bibinfo {author}
  {\bibfnamefont{N.~J.}\ \bibnamefont{van Druten}},\ and\ \bibinfo {author}
  {\bibfnamefont{K.~V.}\ \bibnamefont{Kheruntsyan}},\ }%
  \bibfield{journal}{%
  \Doi{10.1103/PhysRevA.85.031604}{\bibinfo {journal} {Phys. Rev. A}}\ }%
  \textbf{\bibinfo {volume} {85}},\ \bibinfo {pages} {031604(R)} (\bibinfo
  {month} {Mar.}\ \bibinfo {year} {2012}),\
  \url{http://link.aps.org/doi/10.1103/PhysRevA.85.031604}%
  \bibAnnoteFile{NoStop}{Davis:2012hq}%
\bibitem{Anglin:1997cf}%
  \BibitemOpen
  \bibfield{author}{%
  \bibinfo {author} {\bibfnamefont{J.~R.}\ \bibnamefont{Anglin}},\ }%
  \bibfield{journal}{%
  \Doi{10.1103/PhysRevLett.79.6}{\bibinfo {journal} {Phys. Rev. Lett.}}\ }%
  \textbf{\bibinfo {volume} {79}},\ \bibinfo {pages} {6} (\bibinfo {month}
  {Jul.}\ \bibinfo {year} {1997}),\
  \url{http://link.aps.org/doi/10.1103/PhysRevLett.79.6}%
  \bibAnnoteFile{NoStop}{Anglin:1997cf}%
\bibitem{Anglin:1999fn}%
  \BibitemOpen
  \bibfield{author}{%
  \bibinfo {author} {\bibfnamefont{J.~R.}\ \bibnamefont{Anglin}}\ and\ \bibinfo
  {author} {\bibfnamefont{W.~H.}\ \bibnamefont{Zurek}},\ }%
  \bibfield{journal}{%
  \Doi{10.1103/PhysRevLett.83.1707}{\bibinfo {journal} {Phys. Rev. Lett.}}\ }%
  \textbf{\bibinfo {volume} {83}},\ \bibinfo {pages} {1707} (\bibinfo {month}
  {Aug.}\ \bibinfo {year} {1999}),\
  \url{http://link.aps.org/doi/10.1103/PhysRevLett.83.1707}%
  \bibAnnoteFile{NoStop}{Anglin:1999fn}%
\bibitem{TrujilloMartinez:2009gh}%
  \BibitemOpen
  \bibfield{author}{%
  \bibinfo {author} {\bibfnamefont{M.}~\bibnamefont{Trujillo-Martinez}},
  \bibinfo {author} {\bibfnamefont{A.}~\bibnamefont{Posazhennikova}},\ and\
  \bibinfo {author} {\bibfnamefont{J.}~\bibnamefont{Kroha}},\ }%
  \bibfield{journal}{%
  \Doi{10.1103/PhysRevLett.103.105302}{\bibinfo {journal} {Phys. Rev. Lett.}}\
  }%
  \textbf{\bibinfo {volume} {103}},\ \bibinfo {pages} {105302} (\bibinfo
  {month} {Sep.}\ \bibinfo {year} {2009}),\
  \url{http://link.aps.org/doi/10.1103/PhysRevLett.103.105302}%
  \bibAnnoteFile{NoStop}{TrujilloMartinez:2009gh}%
\bibitem{Poletti:2012di}%
  \BibitemOpen
  \bibfield{author}{%
  \bibinfo {author} {\bibfnamefont{D.}~\bibnamefont{Poletti}}, \bibinfo
  {author} {\bibfnamefont{J.-S.}\ \bibnamefont{Bernier}}, \bibinfo {author}
  {\bibfnamefont{A.}~\bibnamefont{Georges}},\ and\ \bibinfo {author}
  {\bibfnamefont{C.}~\bibnamefont{Kollath}},\ }%
  \bibfield{journal}{%
  \Doi{10.1103/PhysRevLett.109.045302}{\bibinfo {journal} {Phys. Rev. Lett.}}\
  }%
  \textbf{\bibinfo {volume} {109}},\ \bibinfo {pages} {045302} (\bibinfo
  {month} {Jul.}\ \bibinfo {year} {2012}),\
  \url{http://link.aps.org/doi/10.1103/PhysRevLett.109.045302}%
  \bibAnnoteFile{NoStop}{Poletti:2012di}%
\bibitem{Kordas:2013gw}%
  \BibitemOpen
  \bibfield{author}{%
  \bibinfo {author} {\bibfnamefont{G.}~\bibnamefont{Kordas}}, \bibinfo {author}
  {\bibfnamefont{S.}~\bibnamefont{Wimberger}},\ and\ \bibinfo {author}
  {\bibfnamefont{D.}~\bibnamefont{Witthaut}},\ }%
  \bibfield{journal}{%
  \Doi{10.1103/PhysRevA.87.043618}{\bibinfo {journal} {Phys. Rev. A}}\ }%
  \textbf{\bibinfo {volume} {87}},\ \bibinfo {pages} {043618} (\bibinfo {month}
  {Apr.}\ \bibinfo {year} {2013}),\
  \url{http://link.aps.org/doi/10.1103/PhysRevA.87.043618}%
  \bibAnnoteFile{NoStop}{Kordas:2013gw}%
\bibitem{Blakie:2008is}%
  \BibitemOpen
  \bibfield{author}{%
  \bibinfo {author} {\bibfnamefont{P.~B.}\ \bibnamefont{Blakie}}, \bibinfo
  {author} {\bibfnamefont{A.~S.}\ \bibnamefont{Bradley}}, \bibinfo {author}
  {\bibfnamefont{M.~J.}\ \bibnamefont{Davis}}, \bibinfo {author}
  {\bibfnamefont{R.~J.}\ \bibnamefont{Ballagh}},\ and\ \bibinfo {author}
  {\bibfnamefont{C.~W.}\ \bibnamefont{Gardiner}},\ }%
  \bibfield{journal}{%
  \Doi{10.1080/00018730802564254}{\bibinfo {journal} {Adv. Phys.}}\ }%
  \textbf{\bibinfo {volume} {57}},\ \bibinfo {pages} {363} (\bibinfo {month}
  {Sep.}\ \bibinfo {year} {2008}),\
  \url{http://www.tandfonline.com/doi/abs/10.1080/00018730802564254}%
  \bibAnnoteFile{NoStop}{Blakie:2008is}%
\bibitem{Gilz:2011jma}%
  \BibitemOpen
  \bibfield{author}{%
  \bibinfo {author} {\bibfnamefont{L.}~\bibnamefont{Gilz}}\ and\ \bibinfo
  {author} {\bibfnamefont{J.~R.}\ \bibnamefont{Anglin}},\ }%
  \bibfield{journal}{%
  \Doi{10.1103/PhysRevLett.107.090601}{\bibinfo {journal} {Phys. Rev. Lett.}}\
  }%
  \textbf{\bibinfo {volume} {107}},\ \bibinfo {pages} {090601} (\bibinfo
  {month} {Aug.}\ \bibinfo {year} {2011}),\
  \url{http://link.aps.org/doi/10.1103/PhysRevLett.107.090601}%
  \bibAnnoteFile{NoStop}{Gilz:2011jma}%
\bibitem{Rooney:2014kc}%
  \BibitemOpen
  \bibfield{author}{%
  \bibinfo {author} {\bibfnamefont{S.~J.}\ \bibnamefont{Rooney}}, \bibinfo
  {author} {\bibfnamefont{P.~B.}\ \bibnamefont{Blakie}},\ and\ \bibinfo
  {author} {\bibfnamefont{A.~S.}\ \bibnamefont{Bradley}},\ }%
  \bibfield{journal}{%
  \Doi{10.1103/PhysRevE.89.013302}{\bibinfo {journal} {Phys. Rev. E}}\ }%
  \textbf{\bibinfo {volume} {89}},\ \bibinfo {pages} {013302} (\bibinfo {month}
  {Jan.}\ \bibinfo {year} {2014}),\
  \url{http://link.aps.org/doi/10.1103/PhysRevE.89.013302}%
  \bibAnnoteFile{NoStop}{Rooney:2014kc}%
\bibitem{Bradley:2014a}%
  \BibitemOpen
  \bibfield{author}{%
  \bibinfo {author} {\bibfnamefont{A.~S.}\ \bibnamefont{Bradley}}\ and\
  \bibinfo {author} {\bibfnamefont{P.~B.}\ \bibnamefont{Blakie}},\ }%
  \bibfield{journal}{%
  \Doi{10.1103/PhysRevA.90.023631}{\bibinfo {journal} {Phys. Rev. A}}\ }%
  \textbf{\bibinfo {volume} {90}},\ \bibinfo {pages} {023631} (\bibinfo {month}
  {Aug.}\ \bibinfo {year} {2014}),\
  \url{http://link.aps.org/doi/10.1103/PhysRevA.90.023631}%
  \bibAnnoteFile{NoStop}{Bradley:2014a}%
\bibitem{Cockburn10a}%
  \BibitemOpen
  \bibfield{author}{%
  \bibinfo {author} {\bibfnamefont{S.~P.}\ \bibnamefont{Cockburn}}, \bibinfo
  {author} {\bibfnamefont{H.~E.}\ \bibnamefont{Nistazakis}}, \bibinfo {author}
  {\bibfnamefont{T.~P.}\ \bibnamefont{Horikis}}, \bibinfo {author}
  {\bibfnamefont{P.~G.}\ \bibnamefont{Kevrekidis}}, \bibinfo {author}
  {\bibfnamefont{N.~P.}\ \bibnamefont{Proukakis}},\ and\ \bibinfo {author}
  {\bibfnamefont{D.~J.}\ \bibnamefont{Frantzeskakis}},\ }%
  \bibfield{journal}{%
  \Doi{10.1103/PhysRevLett.104.174101}{\bibinfo {journal} {Phys. Rev. Lett.}}\
  }%
  \textbf{\bibinfo {volume} {104}},\ \bibinfo {pages} {174101} (\bibinfo
  {month} {Apr.}\ \bibinfo {year} {2010}),\
  \url{http://link.aps.org/doi/10.1103/PhysRevLett.104.174101}%
  \bibAnnoteFile{NoStop}{Cockburn10a}%
\bibitem{Cockburn:2011fa}%
  \BibitemOpen
  \bibfield{author}{%
  \bibinfo {author} {\bibfnamefont{S.~P.}\ \bibnamefont{Cockburn}}, \bibinfo
  {author} {\bibfnamefont{H.~E.}\ \bibnamefont{Nistazakis}}, \bibinfo {author}
  {\bibfnamefont{T.~P.}\ \bibnamefont{Horikis}}, \bibinfo {author}
  {\bibfnamefont{P.~G.}\ \bibnamefont{Kevrekidis}}, \bibinfo {author}
  {\bibfnamefont{N.}~\bibnamefont{P~Proukakis}},\ and\ \bibinfo {author}
  {\bibfnamefont{D.}~\bibnamefont{J~Frantzeskakis}},\ }%
  \bibfield{journal}{%
  \Doi{10.1103/PhysRevA.84.043640}{\bibinfo {journal} {Phys. Rev. A}}\ }%
  \textbf{\bibinfo {volume} {84}},\ \bibinfo {pages} {043640} (\bibinfo {month}
  {Oct.}\ \bibinfo {year} {2011}),\
  \url{http://link.aps.org/doi/10.1103/PhysRevA.84.043640}%
  \bibAnnoteFile{NoStop}{Cockburn:2011fa}%
\bibitem{Su:2013dh}%
  \BibitemOpen
  \bibfield{author}{%
  \bibinfo {author} {\bibfnamefont{S.-W.}\ \bibnamefont{Su}}, \bibinfo {author}
  {\bibfnamefont{S.-C.}\ \bibnamefont{Gou}}, \bibinfo {author}
  {\bibfnamefont{A.~S.}\ \bibnamefont{Bradley}}, \bibinfo {author}
  {\bibfnamefont{O.}~\bibnamefont{Fialko}},\ and\ \bibinfo {author}
  {\bibfnamefont{J.}~\bibnamefont{Brand}},\ }%
  \bibfield{journal}{%
  \Doi{10.1103/PhysRevLett.110.215302}{\bibinfo {journal} {Phys. Rev. Lett.}}\
  }%
  \textbf{\bibinfo {volume} {110}},\ \bibinfo {pages} {215302} (\bibinfo
  {month} {May}\ \bibinfo {year} {2013}),\
  \url{http://link.aps.org/doi/10.1103/PhysRevLett.110.215302}%
  \bibAnnoteFile{NoStop}{Su:2013dh}%
\bibitem{Damski10a}%
  \BibitemOpen
  \bibfield{author}{%
  \bibinfo {author} {\bibfnamefont{B.}~\bibnamefont{Damski}}\ and\ \bibinfo
  {author} {\bibfnamefont{W.~H.}\ \bibnamefont{Zurek}},\ }%
  \bibfield{journal}{%
  \Doi{10.1103/PhysRevLett.104.160404}{\bibinfo {journal} {Phys. Rev. Lett.}}\
  }%
  \textbf{\bibinfo {volume} {104}},\ \bibinfo {pages} {160404} (\bibinfo
  {month} {Apr.}\ \bibinfo {year} {2010}),\
  \url{http://link.aps.org/doi/10.1103/PhysRevLett.104.160404}%
  \bibAnnoteFile{NoStop}{Damski10a}%
\bibitem{Navon:2015jd}%
  \BibitemOpen
  \bibfield{author}{%
  \bibinfo {author} {\bibfnamefont{N.}~\bibnamefont{Navon}}, \bibinfo {author}
  {\bibfnamefont{A.~L.}\ \bibnamefont{Gaunt}}, \bibinfo {author}
  {\bibfnamefont{R.~P.}\ \bibnamefont{Smith}},\ and\ \bibinfo {author}
  {\bibfnamefont{Z.}~\bibnamefont{Hadzibabic}},\ }%
  \bibfield{journal}{%
  \Doi{10.1126/science.1258676}{\bibinfo {journal} {Science}}\ }%
  \textbf{\bibinfo {volume} {347}},\ \bibinfo {pages} {167} (\bibinfo {month}
  {Jan.}\ \bibinfo {year} {2015}),\
  \url{http://www.sciencemag.org/content/347/6218/167.full}%
  \bibAnnoteFile{NoStop}{Navon:2015jd}%
\bibitem{Bradley:2012ih}%
  \BibitemOpen
  \bibfield{author}{%
  \bibinfo {author} {\bibfnamefont{A.~S.}\ \bibnamefont{Bradley}}\ and\
  \bibinfo {author} {\bibfnamefont{B.~P.}\ \bibnamefont{Anderson}},\ }%
  \bibfield{journal}{%
  \Doi{10.1103/PhysRevX.2.041001}{\bibinfo {journal} {Phys. Rev. X}}\ }%
  \textbf{\bibinfo {volume} {2}},\ \bibinfo {pages} {041001} (\bibinfo {month}
  {Oct.}\ \bibinfo {year} {2012}),\
  \url{http://link.aps.org/doi/10.1103/PhysRevX.2.041001}%
  \bibAnnoteFile{NoStop}{Bradley:2012ih}%
\bibitem{Reeves:2013hy}%
  \BibitemOpen
  \bibfield{author}{%
  \bibinfo {author} {\bibfnamefont{M.~T.}\ \bibnamefont{Reeves}}, \bibinfo
  {author} {\bibfnamefont{T.~P.}\ \bibnamefont{Billam}}, \bibinfo {author}
  {\bibfnamefont{B.~P.}\ \bibnamefont{Anderson}},\ and\ \bibinfo {author}
  {\bibfnamefont{A.~S.}\ \bibnamefont{Bradley}},\ }%
  \bibfield{journal}{%
  \Doi{10.1103/PhysRevLett.110.104501}{\bibinfo {journal} {Phys. Rev. Lett.}}\
  }%
  \textbf{\bibinfo {volume} {110}},\ \bibinfo {pages} {104501} (\bibinfo
  {month} {Mar.}\ \bibinfo {year} {2013}),\
  \url{http://link.aps.org/doi/10.1103/PhysRevLett.110.104501}%
  \bibAnnoteFile{NoStop}{Reeves:2013hy}%
\bibitem{Neely:2013ef}%
  \BibitemOpen
  \bibfield{author}{%
  \bibinfo {author} {\bibfnamefont{T.~W.}\ \bibnamefont{Neely}}, \bibinfo
  {author} {\bibfnamefont{A.~S.}\ \bibnamefont{Bradley}}, \bibinfo {author}
  {\bibfnamefont{E.~C.}\ \bibnamefont{Samson}}, \bibinfo {author}
  {\bibfnamefont{S.~J.}\ \bibnamefont{Rooney}}, \bibinfo {author}
  {\bibfnamefont{E.~M.}\ \bibnamefont{Wright}}, \bibinfo {author}
  {\bibfnamefont{K.~J.~H.}\ \bibnamefont{Law}}, \bibinfo {author}
  {\bibfnamefont{R.}~\bibnamefont{Carretero-Gonz{\'a}lez}}, \bibinfo {author}
  {\bibfnamefont{P.~G.}\ \bibnamefont{Kevrekidis}}, \bibinfo {author}
  {\bibfnamefont{M.~J.}\ \bibnamefont{Davis}},\ and\ \bibinfo {author}
  {\bibfnamefont{B.~P.}\ \bibnamefont{Anderson}},\ }%
  \bibfield{journal}{%
  \Doi{10.1103/PhysRevLett.111.235301}{\bibinfo {journal} {Phys. Rev. Lett.}}\
  }%
  \textbf{\bibinfo {volume} {111}},\ \bibinfo {pages} {235301} (\bibinfo
  {month} {Dec.}\ \bibinfo {year} {2013}),\
  \url{http://link.aps.org/doi/10.1103/PhysRevLett.111.235301}%
  \bibAnnoteFile{NoStop}{Neely:2013ef}%
\bibitem{Arnold:2011gy}%
  \BibitemOpen
  \bibfield{author}{%
  \bibinfo {author} {\bibfnamefont{K.~J.}\ \bibnamefont{Arnold}}\ and\ \bibinfo
  {author} {\bibfnamefont{M.~D.}\ \bibnamefont{Barrett}},\ }%
  \bibfield{journal}{%
  \Doi{10.1016/j.optcom.2011.03.008}{\bibinfo {journal} {Opt. Comm.}}\ }%
  \textbf{\bibinfo {volume} {284}},\ \bibinfo {pages} {3288} (\bibinfo {month}
  {Jun.}\ \bibinfo {year} {2011}),\
  \url{http://linkinghub.elsevier.com/retrieve/pii/S0030401811002793}%
  \bibAnnoteFile{NoStop}{Arnold:2011gy}%
\bibitem{Henderson:2009eo}%
  \BibitemOpen
  \bibfield{author}{%
  \bibinfo {author} {\bibfnamefont{K.}~\bibnamefont{Henderson}}, \bibinfo
  {author} {\bibfnamefont{C.}~\bibnamefont{Ryu}}, \bibinfo {author}
  {\bibfnamefont{C.}~\bibnamefont{MacCormick}},\ and\ \bibinfo {author}
  {\bibfnamefont{M.~G.}\ \bibnamefont{Boshier}},\ }%
  \bibfield{journal}{%
  \Doi{10.1088/1367-2630/11/4/043030}{\bibinfo {journal} {New J Phys}}\ }%
  \textbf{\bibinfo {volume} {11}},\ \bibinfo {pages} {043030} (\bibinfo {month}
  {Apr.}\ \bibinfo {year} {2009}),\
  \url{http://stacks.iop.org/1367-2630/11/i=4/a=043030?key=crossref.8564930b7f159cb00bd5536aaaacaf17}%
  \bibAnnoteFile{NoStop}{Henderson:2009eo}%
\bibitem{Gaunt:2013ip}%
  \BibitemOpen
  \bibfield{author}{%
  \bibinfo {author} {\bibfnamefont{A.~L.}\ \bibnamefont{Gaunt}}, \bibinfo
  {author} {\bibfnamefont{T.~F.}\ \bibnamefont{Schmidutz}}, \bibinfo {author}
  {\bibfnamefont{I.}~\bibnamefont{Gotlibovych}}, \bibinfo {author}
  {\bibfnamefont{R.~P.}\ \bibnamefont{Smith}},\ and\ \bibinfo {author}
  {\bibfnamefont{Z.}~\bibnamefont{Hadzibabic}},\ }%
  \bibfield{journal}{%
  \Doi{10.1103/PhysRevLett.110.200406}{\bibinfo {journal} {Phys. Rev. Lett.}}\
  }%
  \textbf{\bibinfo {volume} {110}},\ \bibinfo {pages} {200406} (\bibinfo
  {month} {May}\ \bibinfo {year} {2013}),\
  \url{http://link.aps.org/doi/10.1103/PhysRevLett.110.200406}%
  \bibAnnoteFile{NoStop}{Gaunt:2013ip}%
\bibitem{Steel:1998jr}%
  \BibitemOpen
  \bibfield{author}{%
  \bibinfo {author} {\bibfnamefont{M.~J.}\ \bibnamefont{Steel}}, \bibinfo
  {author} {\bibfnamefont{M.~K.}\ \bibnamefont{Olsen}}, \bibinfo {author}
  {\bibfnamefont{L.~I.}\ \bibnamefont{Plimak}}, \bibinfo {author}
  {\bibfnamefont{P.~D.}\ \bibnamefont{Drummond}}, \bibinfo {author}
  {\bibfnamefont{S.~M.}\ \bibnamefont{Tan}}, \bibinfo {author}
  {\bibfnamefont{M.~J.}\ \bibnamefont{Collett}}, \bibinfo {author}
  {\bibfnamefont{D.~F.}\ \bibnamefont{Walls}},\ and\ \bibinfo {author}
  {\bibfnamefont{R.}~\bibnamefont{Graham}},\ }%
  \bibfield{journal}{%
  \Doi{10.1103/PhysRevA.58.4824}{\bibinfo {journal} {Phys. Rev. A}}\ }%
  \textbf{\bibinfo {volume} {58}},\ \bibinfo {pages} {4824} (\bibinfo {month}
  {Dec.}\ \bibinfo {year} {1998}),\
  \url{http://link.aps.org/doi/10.1103/PhysRevA.58.4824}%
  \bibAnnoteFile{NoStop}{Steel:1998jr}%
\bibitem{Davis2001a}%
  \BibitemOpen
  \bibfield{author}{%
  \bibinfo {author} {\bibfnamefont{M.~J.}\ \bibnamefont{Davis}}, \bibinfo
  {author} {\bibfnamefont{R.~J.}\ \bibnamefont{Ballagh}},\ and\ \bibinfo
  {author} {\bibfnamefont{K.}~\bibnamefont{Burnett}},\ }%
  \bibfield{journal}{%
  \Doi{10.1088/0953-4075/34/22/316}{\bibinfo {journal} {J. Phys. B: At. Mol.
  Opt. Phys.}}\ }%
  \textbf{\bibinfo {volume} {34}},\ \bibinfo {pages} {4487} (\bibinfo {year}
  {2001}),\
  \url{http://stacks.iop.org/0953-4075/34/i=22/a=316?key=crossref.3e1e55ef53a0b448f2c669fbf125447c}%
  \bibAnnoteFile{NoStop}{Davis2001a}%
\bibitem{Wright:2011ey}%
  \BibitemOpen
  \bibfield{author}{%
  \bibinfo {author} {\bibfnamefont{T.~M.}\ \bibnamefont{Wright}}, \bibinfo
  {author} {\bibfnamefont{N.~P.}\ \bibnamefont{Proukakis}},\ and\ \bibinfo
  {author} {\bibfnamefont{M.~J.}\ \bibnamefont{Davis}},\ }%
  \bibfield{journal}{%
  \Doi{10.1103/PhysRevA.84.023608}{\bibinfo {journal} {Phys. Rev. A}}\ }%
  \textbf{\bibinfo {volume} {84}},\ \bibinfo {pages} {023608} (\bibinfo {month}
  {Aug.}\ \bibinfo {year} {2011}),\
  \url{http://link.aps.org/doi/10.1103/PhysRevA.84.023608}%
  \bibAnnoteFile{NoStop}{Wright:2011ey}%
\bibitem{Note1}%
  \BibitemOpen
  \bibinfo {note} {Note that in general $\delta V(\protect \mathbf {r},t)$ in
  (\ref {Ldef}) can be redefined to include the effective potential arising
  from forward scattering with the $I$-region~\cite
  {Gardiner:2003bk,Bradley:2014a}. This term is typically a small correction
  and we do not consider it further here.}%
  \bibAnnoteFile{Stop}{Note1}%
\bibitem{Dalfovo1999}%
  \BibitemOpen
  \bibfield{author}{%
  \bibinfo {author} {\bibfnamefont{F.}~\bibnamefont{Dalfovo}}, \bibinfo
  {author} {\bibfnamefont{S.}~\bibnamefont{Giorgini}}, \bibinfo {author}
  {\bibfnamefont{L.~P.}\ \bibnamefont{Pitaevskii}},\ and\ \bibinfo {author}
  {\bibfnamefont{S.}~\bibnamefont{Stringari}},\ }%
  \bibfield{journal}{%
  \Doi{10.1103/RevModPhys.71.463}{\bibinfo {journal} {Rev. Mod. Phys.}}\ }%
  \textbf{\bibinfo {volume} {71}},\ \bibinfo {pages} {463} (\bibinfo {month}
  {Apr.}\ \bibinfo {year} {1999}),\
  \url{http://link.aps.org/doi/10.1103/RevModPhys.71.463}%
  \bibAnnoteFile{NoStop}{Dalfovo1999}%
\bibitem{Edwards:2012dh}%
  \BibitemOpen
  \bibfield{author}{%
  \bibinfo {author} {\bibfnamefont{M.}~\bibnamefont{Edwards}}, \bibinfo
  {author} {\bibfnamefont{M.}~\bibnamefont{Krygier}}, \bibinfo {author}
  {\bibfnamefont{H.}~\bibnamefont{Seddiqi}}, \bibinfo {author}
  {\bibfnamefont{B.}~\bibnamefont{Benton}},\ and\ \bibinfo {author}
  {\bibfnamefont{C.~W.}\ \bibnamefont{Clark}},\ }%
  \bibfield{journal}{%
  \Doi{10.1103/PhysRevE.86.056710}{\bibinfo {journal} {Phys. Rev. E}}\ }%
  \textbf{\bibinfo {volume} {86}},\ \bibinfo {pages} {056710} (\bibinfo {month}
  {Nov.}\ \bibinfo {year} {2012}),\
  \url{http://link.aps.org/doi/10.1103/PhysRevE.86.056710}%
  \bibAnnoteFile{NoStop}{Edwards:2012dh}%
\bibitem{Petrov:2000cf}%
  \BibitemOpen
  \bibfield{author}{%
  \bibinfo {author} {\bibfnamefont{D.~S.}\ \bibnamefont{Petrov}}, \bibinfo
  {author} {\bibfnamefont{G.~V.}\ \bibnamefont{Shlyapnikov}},\ and\ \bibinfo
  {author} {\bibfnamefont{J.~T.~M.}\ \bibnamefont{Walraven}},\ }%
  \bibfield{journal}{%
  \Doi{10.1103/PhysRevLett.85.3745}{\bibinfo {journal} {Phys. Rev. Lett.}}\ }%
  \textbf{\bibinfo {volume} {85}},\ \bibinfo {pages} {3745} (\bibinfo {month}
  {Oct.}\ \bibinfo {year} {2000}),\
  \url{http://link.aps.org/doi/10.1103/PhysRevLett.85.3745}%
  \bibAnnoteFile{NoStop}{Petrov:2000cf}%
\bibitem{Gilz:2014eo}%
  \BibitemOpen
  \bibfield{author}{%
  \bibinfo {author} {\bibfnamefont{L.}~\bibnamefont{Gilz}}, \bibinfo {author}
  {\bibfnamefont{L.}~\bibnamefont{Rico-P{\'e}rez}},\ and\ \bibinfo {author}
  {\bibfnamefont{J.~R.}\ \bibnamefont{Anglin}},\ }%
  \bibfield{journal}{%
  \Doi{10.1103/PhysRevA.89.052131}{\bibinfo {journal} {Phys. Rev. A}}\ }%
  \textbf{\bibinfo {volume} {89}},\ \bibinfo {pages} {052131} (\bibinfo {month}
  {May}\ \bibinfo {year} {2014}),\
  \url{http://link.aps.org/doi/10.1103/PhysRevA.89.052131}%
  \bibAnnoteFile{NoStop}{Gilz:2014eo}%
\bibitem{Zaremba:1999iu}%
  \BibitemOpen
  \bibfield{author}{%
  \bibinfo {author} {\bibfnamefont{E.}~\bibnamefont{Zaremba}}, \bibinfo
  {author} {\bibfnamefont{T.}~\bibnamefont{Nikuni}},\ and\ \bibinfo {author}
  {\bibfnamefont{A.}~\bibnamefont{Griffin}},\ }%
  \bibfield{journal}{%
  \Doi{10.1023/A:1021846002995}{\bibinfo {journal} {J. Low Temp. Phys.}}\ }%
  \textbf{\bibinfo {volume} {116}},\ \bibinfo {pages} {277} (\bibinfo {year}
  {1999}),\ \url{http://link.springer.com/10.1023/A:1021846002995}%
  \bibAnnoteFile{NoStop}{Zaremba:1999iu}%
\bibitem{Stoof:1999tz}%
  \BibitemOpen
  \bibfield{author}{%
  \bibinfo {author} {\bibfnamefont{H.~T.~C.}\ \bibnamefont{Stoof}},\ }%
  \bibfield{journal}{%
  \Doi{10.1023/A:1021897703053}{\bibinfo {journal} {J. Low Temp. Phys.}}\ }%
  \textbf{\bibinfo {volume} {114}},\ \bibinfo {pages} {11} (\bibinfo {month}
  {Jan.}\ \bibinfo {year} {1999}),\
  \url{http://link.springer.com/10.1023/A:1021897703053}%
  \bibAnnoteFile{NoStop}{Stoof:1999tz}%
\bibitem{Proukakis:2008eo}%
  \BibitemOpen
  \bibfield{author}{%
  \bibinfo {author} {\bibfnamefont{N.~P.}\ \bibnamefont{Proukakis}}\ and\
  \bibinfo {author} {\bibfnamefont{B.}~\bibnamefont{Jackson}},\ }%
  \bibfield{journal}{%
  \Doi{10.1088/0953-4075/41/20/203002}{\bibinfo {journal} {J. Phys. B: At. Mol.
  Opt. Phys.}}\ }%
  \textbf{\bibinfo {volume} {41}},\ \bibinfo {pages} {203002} (\bibinfo {year}
  {2008}),\ \url{http://www.iop.org/EJ/abstract/0953-4075/41/20/203002/}%
  \bibAnnoteFile{NoStop}{Proukakis:2008eo}%
\bibitem{Gardiner:2007gj}%
  \BibitemOpen
  \bibfield{author}{%
  \bibinfo {author} {\bibfnamefont{S.~A.}\ \bibnamefont{Gardiner}}\ and\
  \bibinfo {author} {\bibfnamefont{S.~A.}\ \bibnamefont{Morgan}},\ }%
  \bibfield{journal}{%
  \Doi{10.1103/PhysRevA.75.043621}{\bibinfo {journal} {Phys. Rev. A}}\ }%
  \textbf{\bibinfo {volume} {75}},\ \bibinfo {pages} {043621} (\bibinfo {year}
  {2007}),\
  \url{http://scitation.aip.org/getabs/servlet/GetabsServlet?prog=normal&id=PLRAAN000075000004043621000001&idtype=cvips&gifs=yes}%
  \bibAnnoteFile{NoStop}{Gardiner:2007gj}%
\bibitem{Billam:2013kb}%
  \BibitemOpen
  \bibfield{author}{%
  \bibinfo {author} {\bibfnamefont{T.~P.}\ \bibnamefont{Billam}}, \bibinfo
  {author} {\bibfnamefont{P.}~\bibnamefont{Mason}},\ and\ \bibinfo {author}
  {\bibfnamefont{S.~A.}\ \bibnamefont{Gardiner}},\ }%
  \bibfield{journal}{%
  \Doi{10.1103/PhysRevA.87.033628}{\bibinfo {journal} {Phys. Rev. A}}\ }%
  \textbf{\bibinfo {volume} {87}},\ \bibinfo {pages} {033628} (\bibinfo {month}
  {Mar.}\ \bibinfo {year} {2013}),\
  \url{http://link.aps.org/doi/10.1103/PhysRevA.87.033628}%
  \bibAnnoteFile{NoStop}{Billam:2013kb}%
\bibitem{Mason:2014dd}%
  \BibitemOpen
  \bibfield{author}{%
  \bibinfo {author} {\bibfnamefont{P.}~\bibnamefont{Mason}}\ and\ \bibinfo
  {author} {\bibfnamefont{S.~A.}\ \bibnamefont{Gardiner}},\ }%
  \bibfield{journal}{%
  \Doi{10.1103/PhysRevA.89.043617}{\bibinfo {journal} {Phys. Rev. A}}\ }%
  \textbf{\bibinfo {volume} {89}},\ \bibinfo {pages} {043617} (\bibinfo {month}
  {Apr.}\ \bibinfo {year} {2014}),\
  \url{http://link.aps.org/doi/10.1103/PhysRevA.89.043617}%
  \bibAnnoteFile{NoStop}{Mason:2014dd}%
\bibitem{Bijlsma:2000hv}%
  \BibitemOpen
  \bibfield{author}{%
  \bibinfo {author} {\bibfnamefont{M.~J.}\ \bibnamefont{Bijlsma}}, \bibinfo
  {author} {\bibfnamefont{E.}~\bibnamefont{Zaremba}},\ and\ \bibinfo {author}
  {\bibfnamefont{H.~T.~C.}\ \bibnamefont{Stoof}},\ }%
  \bibfield{journal}{%
  \Doi{10.1103/PhysRevA.62.063609}{\bibinfo {journal} {Phys. Rev. A}}\ }%
  \textbf{\bibinfo {volume} {62}},\ \bibinfo {pages} {063609} (\bibinfo {year}
  {2000}),\
  \url{http://journals.aps.org/pra/abstract/10.1103/PhysRevA.62.063609}%
  \bibAnnoteFile{NoStop}{Bijlsma:2000hv}%
\bibitem{Duine:2001iu}%
  \BibitemOpen
  \bibfield{author}{%
  \bibinfo {author} {\bibfnamefont{R.~A.}\ \bibnamefont{Duine}}\ and\ \bibinfo
  {author} {\bibfnamefont{H.~T.~C.}\ \bibnamefont{Stoof}},\ }%
  \bibfield{journal}{%
  \Doi{10.1103/PhysRevA.65.013603}{\bibinfo {journal} {Phys. Rev. A}}\ }%
  \textbf{\bibinfo {volume} {65}},\ \bibinfo {pages} {25} (\bibinfo {month}
  {Dec.}\ \bibinfo {year} {2001})%
  \bibAnnoteFile{NoStop}{Duine:2001iu}%
\bibitem{Stoof:2001wk}%
  \BibitemOpen
  \bibfield{author}{%
  \bibinfo {author} {\bibfnamefont{H.~T.~C.}\ \bibnamefont{Stoof}}\ and\
  \bibinfo {author} {\bibfnamefont{M.~J.}\ \bibnamefont{Bijlsma}}\ }%
  \textbf{\bibinfo {volume} {124}},\ \bibinfo {pages} {431} (\bibinfo {year}
  {2001})%
  \bibAnnoteFile{NoStop}{Stoof:2001wk}%
\bibitem{Das:2012ki}%
  \BibitemOpen
  \bibfield{author}{%
  \bibinfo {author} {\bibfnamefont{A.}~\bibnamefont{Das}}, \bibinfo {author}
  {\bibfnamefont{J.}~\bibnamefont{Sabbatini}},\ and\ \bibinfo {author}
  {\bibfnamefont{W.~H.}\ \bibnamefont{Zurek}},\ }%
  \bibfield{journal}{%
  \Doi{10.1038/srep00352}{\bibinfo {journal} {Sci. Rep.}}\ }%
  \textbf{\bibinfo {volume} {2}},\ \bibinfo {pages} {352} (\bibinfo {month}
  {Apr.}\ \bibinfo {year} {2012}),\
  \url{http://www.nature.com/doifinder/10.1038/srep00352}%
  \bibAnnoteFile{NoStop}{Das:2012ki}%
\bibitem{Cockburn:2011kw}%
  \BibitemOpen
  \bibfield{author}{%
  \bibinfo {author} {\bibfnamefont{S.~P.}\ \bibnamefont{Cockburn}}, \bibinfo
  {author} {\bibfnamefont{A.}~\bibnamefont{Negretti}}, \bibinfo {author}
  {\bibfnamefont{N.~P.}\ \bibnamefont{Proukakis}},\ and\ \bibinfo {author}
  {\bibfnamefont{C.}~\bibnamefont{Henkel}},\ }%
  \bibfield{journal}{%
  \Doi{10.1103/PhysRevA.83.043619}{\bibinfo {journal} {Phys. Rev. A}}\ }%
  \textbf{\bibinfo {volume} {83}},\ \bibinfo {pages} {043619} (\bibinfo {month}
  {Apr.}\ \bibinfo {year} {2011}),\
  \url{http://link.aps.org/doi/10.1103/PhysRevA.83.043619}%
  \bibAnnoteFile{NoStop}{Cockburn:2011kw}%
\bibitem{Cockburn:2012gc}%
  \BibitemOpen
  \bibfield{author}{%
  \bibinfo {author} {\bibfnamefont{S.~P.}\ \bibnamefont{Cockburn}}\ and\
  \bibinfo {author} {\bibfnamefont{N.~P.}\ \bibnamefont{Proukakis}},\ }%
  \bibfield{journal}{%
  \Doi{10.1103/PhysRevA.86.033610}{\bibinfo {journal} {Phys. Rev. A}}\ }%
  \textbf{\bibinfo {volume} {86}},\ \bibinfo {pages} {033610} (\bibinfo {month}
  {Sep.}\ \bibinfo {year} {2012}),\
  \url{http://link.aps.org/doi/10.1103/PhysRevA.86.033610}%
  \bibAnnoteFile{NoStop}{Cockburn:2012gc}%
\bibitem{Cockburn09a}%
  \BibitemOpen
  \bibfield{author}{%
  \bibinfo {author} {\bibfnamefont{S.~P.}\ \bibnamefont{Cockburn}}\ and\
  \bibinfo {author} {\bibfnamefont{N.~P.}\ \bibnamefont{Proukakis}},\ }%
  \bibfield{journal}{%
  \Doi{10.1134/S1054660X09040057}{\bibinfo {journal} {Laser Phys.}}\ }%
  \textbf{\bibinfo {volume} {19}},\ \bibinfo {pages} {558} (\bibinfo {month}
  {Apr.}\ \bibinfo {year} {2009}),\
  \url{http://link.springer.com/10.1134/S1054660X09040057}%
  \bibAnnoteFile{NoStop}{Cockburn09a}%
\bibitem{Su:2012jp}%
  \BibitemOpen
  \bibfield{author}{%
  \bibinfo {author} {\bibfnamefont{S.~W.}\ \bibnamefont{Su}}, \bibinfo {author}
  {\bibfnamefont{I.~K.}\ \bibnamefont{Liu}}, \bibinfo {author}
  {\bibfnamefont{Y.~C.}\ \bibnamefont{Tsai}}, \bibinfo {author}
  {\bibfnamefont{W.~M.}\ \bibnamefont{Liu}},\ and\ \bibinfo {author}
  {\bibfnamefont{S.~C.}\ \bibnamefont{Gou}},\ }%
  \bibfield{journal}{%
  \Doi{10.1103/PhysRevA.86.023601}{\bibinfo {journal} {Phys. Rev. A}}\ }%
  \textbf{\bibinfo {volume} {86}},\ \bibinfo {pages} {023601} (\bibinfo {month}
  {Aug.}\ \bibinfo {year} {2012}),\
  \url{http://link.aps.org/doi/10.1103/PhysRevA.86.023601}%
  \bibAnnoteFile{NoStop}{Su:2012jp}%
\end{thebibliography}
%

\end{document}